\documentclass[11pt,twoside]{article}

\usepackage{graphicx}
\usepackage{amssymb}
\usepackage{amsmath}

\usepackage{natbib}

\newcommand{\bk}{{\bf k}}

\newcommand{\br}{{\bf r}}

\newcommand{\bA}{{\bf A}}
\newcommand{\bB}{{\bf B}}
\newcommand{\bmu}{{\boldsymbol \mu}}

\newcommand{\kB}{k_{\mathrm{B}}}
\newcommand{\kF}{k_{\mathrm{F}}}
\newcommand{\nF}{n_{\mathrm{F}}}
\newcommand{\Green}{{G}}

\begin{document}

\begin{center}
{\LARGE The Gross-Pitaevskii Equations and Beyond for Inhomogeneous
Condensed Bosons}

\bigskip

G. G. N. Angilella$^1$,
S. Bartalini$^2$,
F. S. Cataliotti$^{1,2,3}$,\\
I. Herrera$^{1,2}$,
N. H. March$^{4,5}$,
R. Pucci$^1$

\bigskip

$^1$Dipartimento di Fisica e Astronomia, Universit\`a di Catania,\\
and Lab. MATIS-INFM, and CNISM, UdR Catania, and INFN, Sez. Catania,\\
Via S. Sofia, 64, I-95123 Catania, Italy.\\

\medskip

$^2$LENS, Dipartimento di Fisica, Universit\`a di Firenze,\\
and Istituto Nazionale per la Fisica della Materia, UdR Firenze,\\
Via Nello Carrara, 1, I-50019 Sesto Fiorentino (FI), Italy.\\

\medskip

$^3$Scuola Superiore di Catania, Universit\`a di Catania, Catania,
   Italy.\\

\medskip

$^4$Department of Physics, University of Antwerp,\\
Groenenborgerlaan 171, B-2020 Antwerp, Belgium.\\

\medskip

$^5$Oxford University, Oxford, England.

\end{center}

\begin{abstract}
A simple derivation of the static Gross-Pitaevskii (GP) equation is
   given from an energy variational principle.
The result is then generalized heuristically to the time-dependent GP
   form.
With this as background, a number of different experimental areas
   explored very recently are reviewed, in each case contact being
   established between the measurements and the predictions of the GP
   equations.
The various limitations of these equations as used on dilute
   inhomogeneous condensed Boson atomic gases are then summarized,
   reference also being made to the fact that there is no many-body
   wave function underlying the GP formulation.
This then leads into a discussion of a recently proposed integral
   equation, derived by taking the Bogoliubov-de~Gennes equation as
   starting point.
Some limitations of the static GP differential equation are thereby
   removed, though it is a matter of further study to determine
   whether a correlated wave function exists as underpinning for the
   integral equation formulation.
\end{abstract}




\section{Introduction}

The purpose of this Chapter is to review recent progress in both
   theory and experiment in the area of inhomogeneous condensed
   bosons.
Both statics and dynamics will be referred to in the course of this
   review.
However, the field cited above is now vast, and therefore we shall
   select, in both experiment and theory, the areas in which our own
   contribution lie.

With the above brief outline, we next emphasize that, since the
   Gross-Pitaevskii (GP) equations, to be given explicitly in
   Section~\ref{sec:GP} below, have been very valuable in providing a
   theoretical framework, albeit quite approximate, for the
   interpretation of an extensive body of experimental data, we shall
   take the GP equations as a focus for the present Chapter.
However, to avoid repetition, we cite here two major reviews which
   should be consulted by the reader who requires more technical
   background to the basic theoretical arguments, and application
   techniques, underlying the GP equations.

The first of these is by \citet{Dalfovo:99}.
This provides an authorative review of the phenomenon of Bose-Einstein
   condensation (BEC) of dilute gases in harmonic traps from a theoretical
   perspective.
For the ideal Bose gas, \citeauthor{Dalfovo:99} treat
\emph{(i)} finite size effects,
\emph{(ii)} the role of dimensionality, and
\emph{(iii)} the thermodynamic limit of trapped Bosons at finite
   temperature.
In addition, again for the ideal Bose gas, nonharmonic traps and
   adiabatic transformations are also treated.
So in the present Chapter we shall take their treatment of the ideal
   Bose gas in traps as assumed background.

The second of the reviews referred to above is of very different
   character, but again provides invaluable background for this
   present Chapter.
Thus the focus of the review by \citet{Minguzzi:04} is on numerical
   methods in use for atomic quantum gases with application to
   Bose-Einstein condensates, our dominant focus in this Chapter, but
   also \citeauthor{Minguzzi:04} cover ultracold Fermions.
Thus the reader interested specifically in numerical techniques used
   to solve practical problems in inhomogeneous assemblies of
   condensed Bosons should consult the extensive treatment of
   \citet{Minguzzi:04}.

With this background, the outline of the present Chapter is as
   follows.
In Section~\ref{sec:GP} both static and dynamic GP equations are set
   up by what we consider to be the simplest physical considerations.
One example to illustrate the time-dependent GP Eq.~\eqref{eq:GPt}
   below, which will be summarized also in Section~\ref{sec:GP}, is drop
   emission from an optical lattice under gravity
   \citep{Anderson:98e,Chiofalo:99}, while a somewhat generalized
   static GP equation is utilized in Fig.~\ref{fig:Castin}, which
   shows a vortex array for a rotating Bose-Einstein condensate.
These two examples lead into Section~\ref{sec:expt}, in which selected
   experiments are both described and also brought into contact with
   the GP equations.
In particular, we shall focus on optical lattices, atom chips and
   magnetic microtraps, and the realization of Josephson Junction
   arrays of Bose-Einstein condensates \citep{Cataliotti:01}.
We will also briefly survey the dynamics of a BEC expanding in a
   moving 1D optical lattice, and some experiments beyond the GP
   equations.
Section~\ref{sec:limitations} is then a very brief discussion, following
   the arguments of \citet{Leggett:03}, of some limitations of the GP
   equations exposed by first-principles arguments.
This leads into Sections~\ref{sec:BdG} and \ref{sec:integral}, in
   which an integral equation transcending the static GP
   Eq.~\eqref{eq:GPs} is set up, following the proposal by
   \citet{Angilella:04d}.
The Chapter concludes with a brief summary and some selected proposals
   for directions in which further studies should prove fruitful
   (Section~\ref{sec:summary}).

\section{Gross-Pitaevskii (GP) equations}
\label{sec:GP}

The equations which provide a focus for the present review were
   proposed independently by \citet{Gross:61,Gross:63} and by
   \citet{Pitaevskii:61}.
However, it is fair to say that the mean-field description of a dilute
   Bose gas goes back at least to \citet{Bogoliubov:47}.
The central point in his study was in separating out the condensate
   contribution.
The generalization of the Bogoliubov approach to embrace the
   situations where there is both inhomogeneity and time-dependence is
   the aim of this section of the present review.

\subsection{Static GP equation}

The natural starting point then is to treat first the static, or
   time-independent case, for inhomogeneous condensed Bosons.
This will lead to the so-called static GP equation.
The approach we shall adopt below
   \citep[see also][]{Dalfovo:99,Minguzzi:04} is to start from variational
   minimization, for a Bose-condensed gas in a three-dimensional trap
   at zero temperature, of the energy functional $E[\Phi]$ given by
\begin{equation}
E[\Phi] = \int d\br \left[ \frac{\hbar^2}{2m} | \nabla \Phi |^2 +
   V_{\mathrm{ext}} (\br) |\Phi|^2 + \frac{1}{2} g |\Phi|^4 \right].
\label{eq:GL}
\end{equation}
In this energy functional, $|\Phi(\br)|^2$ is the inhomogeneous
   density profile, the so-called `condensate wave function',
   $\Phi(\br)$ playing the role of the order parameter of the Boson
   assembly.
The first term on the right-hand-side (RHS) of Eq.~\eqref{eq:GL}
   represents the kinetic energy of the condensate while the second is
   the potential energy due to the external confining potential
   $V_{\mathrm{ext}} (\br)$.
The final term on the RHS of Eq.~\eqref{eq:GL} requires rather more
   discussion.
It is appropriate in a dilute and cold gas as a representation of the
   atom-atom interaction, say in general $U(\br -\br^\prime )$, since
   in this case only binary collisions at low energy prove to have
   relevance.
These collisions can be characterized, in fact, by a single parameter
   $a$, which denotes the $s$-wave scattering length, independently of
   the finer details of $U(\br -\br^\prime )$.
As \citet{Dalfovo:99} stress, this permits one to
   adopt an effective interaction given by
\begin{equation}
U(\br -\br^\prime ) = g \delta (\br -\br^\prime )
\label{eq:gdelta}
\end{equation}
\citep[see also][]{Parkins:98,Leggett:01}.
Here the coupling constant $g$ is given in terms of the $s$-wave
   scattering length $a$ in three dimensions and atomic mass $m$ by
\begin{equation}
g = \frac{4\pi\hbar^2 a}{m} .
\label{eq:scatt}
\end{equation}
It has been emphasized that $g$ is a parameter depending on particle density
   \citep{Pieri:00} as well as on dimensionality \citep[][and
   references therein]{Cherny:04}.
In particular, its dependence on density has been
   discussed by \citet{Pieri:00} in connection with the crossover from
   wek-coupling BCS superconductivity to strong-coupling BE condensation
   \citep[][see also \cite{Randeria:95} for a review]{Nozieres:85}.

Returning then to the functional Eq.~\eqref{eq:GL}, it is to be noted
   that for repulsive interactions corresponding to $g>0$, the
   functional is convex and the minimum corresponds to the stable
   ground state.
It should be emphasized, for the other case when $g<0$, that the
   ground state exists only at weak coupling for a limited number of
   trapped Bosons, so long as the zero-point energy is able to balance
   the effect of attractions and prevent collapse.

With this short introduction to the assumed energy functional
   $E[\Phi]$ set out in Eq.~\eqref{eq:GL}, it is then a
   straightforward matter to perform the minimization with respect to
   the order parameter $\Phi(\br)$.
One is then led to the (of course approximate) static GP differential
   equation having the form of a non-linear Schr\"odinger equation,
   namely
\begin{equation}
\left[ -\frac{\hbar^2}{2m} \nabla^2 + V_{\mathrm{ext}} (\br) + g
   |\Phi(\br)|^2 \right] \Phi(\br) = \mu \Phi(\br),
\label{eq:GPs}
\end{equation}
a condensate at equilibrium being at chemical potential $\mu$.

\begin{figure}[t]
\centering
\includegraphics[width=0.9\columnwidth]{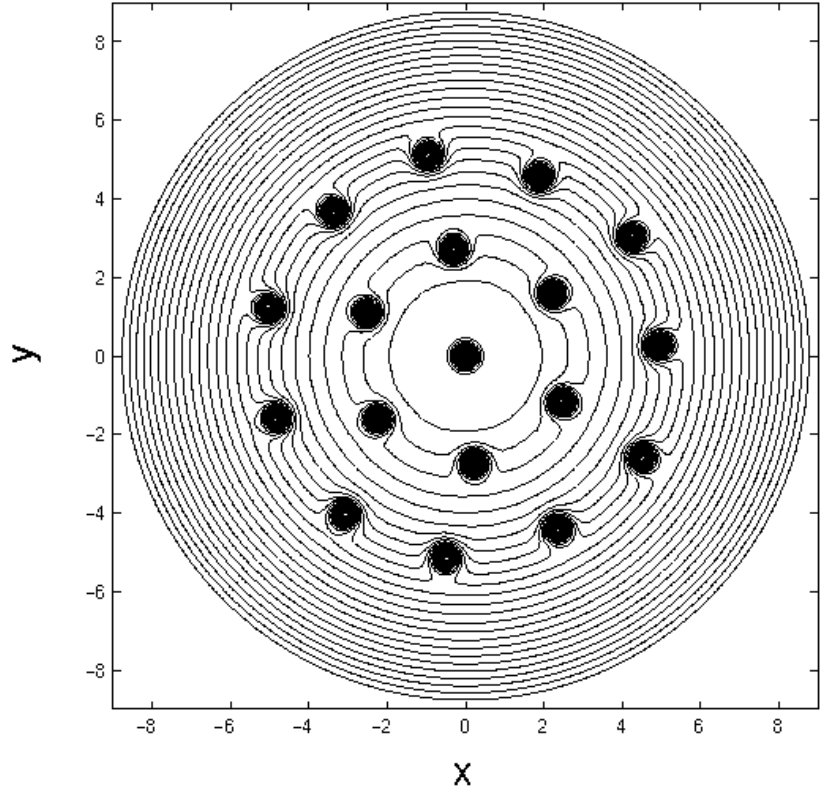}
\caption{Shows vortex array as exhibited by numerical solution of the
   static Gross-Pitaevskii Eq.~(\protect\ref{eq:GPs}) for a rotating
   Bose-Einstein condensate.
Redrawn from \protect\citet{Castin:99}.}
\label{fig:Castin}
\end{figure}

While we shall consider rather fully below a variety of recent
   experimental data obtained by some of the present authors, let us briefly
   summarize at this point just one application of the static GP
   equation, Eq.~\eqref{eq:GPs}.
The example below follows the work of \citet{Castin:99} who treated a
   vortex array.
This they obtained by numerical solution of the two-dimensional
   differential equation for a rotating Bose-Einstein condensate (BEC)
   under tight vertical confinement.
In this rotating condensate, described in the rotating frame by an
   appropriate static GP equation, one needs to incorporate an
   inertial term.
Then Eq.~\eqref{eq:GPs} can be extended to the form
\begin{equation}
\left[ -\frac{\hbar^2}{2m} \nabla^2 + V_{\mathrm{ext}} (\br) + g
   |\Phi(\br)|^2 +\Omega L_z \right] \Phi(\br) = \mu \Phi(\br),
\label{eq:GPr}
\end{equation}
$\Omega$ denoting the rotation frequency and $L_z$ the angular
   momentum component along the rotation axis.
As emphasized in the study of \citet{Castin:99}, the term $\Omega L_z$
   in Eq.~\eqref{eq:GPr} is not diagonal in position or in momentum
   space, and this implies that numerical solution of
   Eq.~\eqref{eq:GPr} requires special care.

Starting with different trial states, a single-vortex solution, as
   well as those describing multi-vortex configurations, can be
   exhibited.
In this latter form of solution, the vortices display a tendency to
   arrange themselves into a triangular geometry, as shown in
   Fig.~\ref{fig:Castin} which is redrawn from \citet{Castin:99}.
As stressed by \citet{Minguzzi:04}, a significant result which has
   emerged from solving the static GP Eq.~\eqref{eq:GPr} in a
   cigar-shaped trap is that a vortex line can be bent, as shown in
   Fig.~17 of \citet{Minguzzi:04}, following the study of
   \citet{Garcia-Ripoll:01}.
This finding is in general agreement with experimental observations.

\subsection{Inhomogeneous Bose superfluids: dynamics using the
   time-dependent GP equation}

Let us proceed to treat in a somewhat parallel manner the dynamics of
   inhomogeneous trapped superfluid Boson gases.
One can make a heuristic generalization of the non-linear
   Schr\"odinger equation having the form of the static GP
   Eq.~\eqref{eq:GPs} to describe the now time-dependent condensate
   wave function $\Phi(\br,t)$, to read
\begin{equation}
\left[ - \frac{\hbar^2}{2m} \nabla^2 + V_{\mathrm{ext}} (\br) + g
   |\Phi(\br,t)|^2 \right] \Phi(\br,t) = i \hbar
   \frac{\partial}{\partial t} \Phi(\br,t)
\label{eq:GPt}
\end{equation}
which, of course, reduces to the static Eq.~\eqref{eq:GPs} on making
   the substitution $\Phi(\br,t) = \Phi(\br) \exp ( - i \mu t /\hbar
   )$.
Eq.~\eqref{eq:GPt}, also given by \citet{Gross:61,Gross:63} and
   \citet{Pitaevskii:61}, will be discussed in some detail below, in
   order to gain further insight into the realm of validity.
Following \citet{Dalfovo:99}, let us write down the many-body
   Hamiltonian $\hat{H}$, in second quantization, which describes $N$
   interacting Bosons confined again by an external potential
   $V_{\mathrm{ext}}$.
This reads:
\begin{eqnarray}
\hat{H} &=& \int d\br \hat{\Psi}^\dag (\br) \left[ -
   \frac{\hbar^2}{2m} \nabla^2 + V_{\mathrm{ext}} (\br)\right] \hat{\Psi}
   (\br) \nonumber \\
&&+ \frac{1}{2} \int\int d\br d\br^\prime
\hat{\Psi}^\dag (\br) \hat{\Psi}^\dag (\br^\prime ) U(\br - \br^\prime
   ) \hat{\Psi} (\br^\prime ) \hat{\Psi} (\br),
\label{eq:HB}
\end{eqnarray}
where $U(\br-\br^\prime )$ is the two-body interatomic potential
   already referred to.
In Eq.~\eqref{eq:HB}, $\hat{\Psi} (\br)$ and $\hat{\Psi}^\dag (\br)$ are the
   Bosonic field operators in the Schr\"odinger representation ($t=0$,
   say).
Then, it follows that the evolution equation for the operator
   $\hat{\Psi} (\br,t)$ takes the form
\begin{equation}
i\hbar \frac{\partial}{\partial t} \hat{\Psi} (\br,t) =
\left[ - \frac{\hbar^2}{2m} \nabla^2 + V_{\mathrm{ext}} (\br)
+ \int d\br^\prime \hat{\Psi}^\dag (\br^\prime , t) U(\br - \br^\prime
   ) \hat{\Psi} (\br^\prime ,t)\right] \hat{\Psi} (\br,t).
\end{equation}
For a very dilute and fully Bose-Einstein condensed gas at $T=0$, it
   is assumed that the field operator can be replaced by the classical
   field $\Phi(\br,t)$.
Taking again the description of $U(\br - \br^\prime )$ in
   Eq.~\eqref{eq:gdelta} with $g$ related to the $s$-wave scattering
   length $a$ by Eq.~\eqref{eq:scatt}, one is led back to the
   time-dependent GP Eq.~\eqref{eq:GPt}.

It is known that the time-dependent GP equation, because it omits
   dissipation, is not appropriate to deal with the dynamics of a
   confined Bose gas when quantum depletion or thermal excitations
   play a significant role.
As examples of such situations, one may cite condensate formation and
   decay, phase coherence, damping of collective motions and
   excitations from a non--mean-field ground state.
Nevertheless, the time-dependent GP Eq.~\eqref{eq:GPt} has been found
   valuable in treating a variety of dynamical processes in condensate
   clouds.
One may cite, as examples, collective-mode frequencies in harmonic or
   optical-lattice confinement, interference phenomena, dynamics of
   vortices, propagation of solitons, shock-wave dynamics, four-wave
   mixing, atom-laser output, as well as expansion of a rotating
   condensate.

While numerous examples of the use of both GP Eqs.~\eqref{eq:GPs} and
   \eqref{eq:GPt} will be given below in Sec.~\ref{sec:expt}, one
   immediate illustration to show the usefulness of Eq.~\eqref{eq:GPt}
   will be taken from the work of \citet{Chiofalo:99}.
Their extensive numerical study of drop emission from an optical
   lattice under gravity was motivated by the experiment of
   \citet{Anderson:98e}.
In their experiment, an almost pure BEC was poured from a
   magneto-optic trap into a vertical optical lattice which was
   produced by a detuned standing wave of light from the
   counter-propagating laser beams.
The gravitational force tilts the lattice potential and drives
   tunnelling from well states to the continuum.
Interference taking place between coherent blobs of condensate at
   different lattice sites leads to the appearance of falling drops,
   which can be regarded as coherent matter-wave pulses by analogy
   with a mode-locked photon laser.
Analogous to the laser cavity is the Brillouin zone in momentum space,
   so that the modulation interval for the pulsed emission of drops
   equals the period of Bloch oscillations.
The time interval between successive drops was $1.1$~ms in both the
   experiment and the numerical study, which is the period of Bloch
   oscillation for a quasi-particle (the whole coherent condensate)
   driven by the constant force of gravity through a periodic array of
   potential wells having a lattice period of one-half of the laser
   wavelength $\lambda$.

\begin{figure}[t]
\centering
\includegraphics[width=0.7\textwidth]{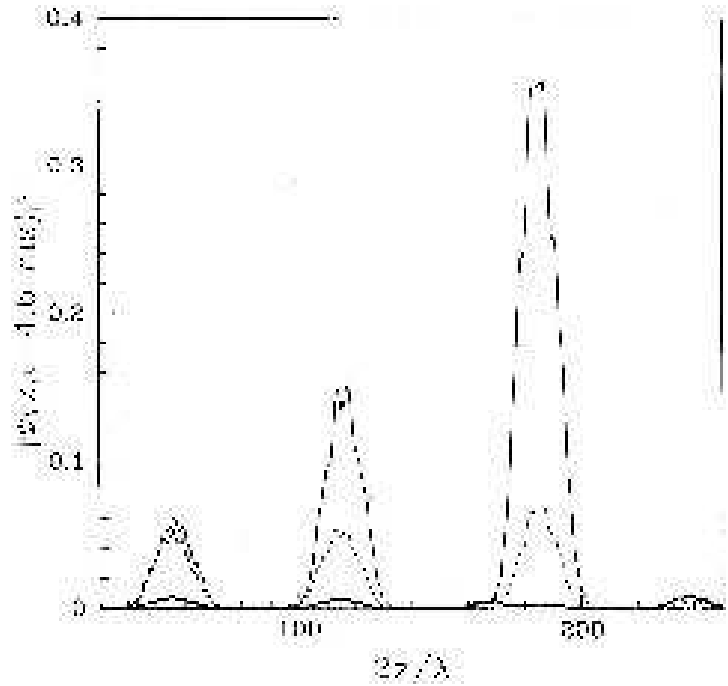}
\caption{Plots shown relate to condensate drop emission from an
   optical lattice under gravity.
Actually shown is the density profile of the condensate drops after
   $4.6$~ms as a function of a (suitably scaled) distance $z$ in the
   non-interacting limit.
The central condensate around $z=0$ has been substracted out.
Different curves plotted are characterized by increasing well depths,
   going from shallow (intermediate thickness curve) to deep (thickest
   curve).
Redrawn from \protect\citet{Chiofalo:99}.}
\label{fig:Chiofalo}
\end{figure}

Primary attention was paid by \citet{Chiofalo:99} in their numerical
   investigations to the part played by atomic interactions in this
   example of coherent transport.
In agreement with band-structure theory of transport in periodic
   structures, the period of drop emission depends only on the
   intensity of the drive and the lattice spacing.
The size and shape of the emitted drops are instead determined by the
   magnitude of the lattice barrier and the interatomic forces.
In Fig.~\ref{fig:Chiofalo}, redrawn from \citet{Chiofalo:99}, the
   density profile of the drops emitted from a non-interacting BEC
   after $4.6$~ms is depicted as a function of a suitably scaled
   distance $z$.
The main condensate around $z=0$ has been subtracted, the curves being
   characterized by values of the lattice barrier height.
In the experiment, the latter is determined by the intensity of the
   optical beams.

When the constant drive is supplemented by a monochromatic oscillating
   drive, it turns out that the equations governing this form of
   coherent BEC transport in the linear non-dissipative regime can be
   mapped onto those for the ac~superconducting current flowing across
   a weak-link Josephson junction \citep{Chiofalo:01}.
The voltage drop across the junction is, in essence, replaced by the
   product of the constant component of the force times the lattice
   spacing, and the frequency of the Bloch oscillations is in
   resonance with integer multiples of the oscillating drive
   frequency.
\citet{Burger:01} have carried out the relevant experiment on a BEC
   confined in a magnetic trap plus an optical lattice, the
   oscillating force being generated by a rapid shift of the centre of
   the trap.

With this introduction to both static and time-dependent GP
   differential equations given in Eqs.~\eqref{eq:GPs} and
   \eqref{eq:GPt} respectively, we turn next to discuss a variety of
   recent experiments by some of the present authors, and the
   relevance of these GP equations to their interpretation.

\section{Selected experiments brought into contact with GP equations}
\label{sec:expt}

The first experimental realization of a Bose--Einstein condensate
(BEC) in a dilute gas of rubidium atoms \citep[see][as a general
   reference]{Varenna:98} has marked the birth of a completely new
   field of optics, that of coherent matter wave beams.
Indeed, it was immediately recognized that a BEC is in very many
   aspects the analogue of an optical laser for atoms.
After the first pioneering experiment at MIT where a pulsed atom laser
   was created by simply switching off the confining magnetic
   potential \citep{Mewes:97}, other more refined experiments followed.
The first continuous wave \emph{atom laser} was realized in Munich by
   coupling atoms out of a magnetic trap using a radio frequency field
   \citep{Bloch:99}, while in Yale a BEC was loaded into a vertical
   standing wave to produce the analogue of a mode-locked laser
   \citep{Anderson:98e}.
Coherent matter wave beams were soon shown to produce interference
   fringes by \citet{Andrews:97}, and double slit and multiple beam
   interferometers were soon realized \citep{Fort:01,Pedri:01}.
On the other hand atoms, at variance with photons, interact even in
   the dilute gas limit.
The nonlinearity introduced by atomic interactions in coherent matter
   waves propagation through vacuum was readily shown to be the same
   of the third order susceptibility for electromagnetic waves
   propagating through a nonlinear crystal.
It was soon demonstrated that it was indeed possible to observe matter
   wave amplification \citep{Kozuma:99,Inouye:99} and four wave mixing
   \citep{Deng:99}.
More recently, the intrinsic nonlinearity of condensates was
   successfully exploited in the observation of dark
   \citep{Burger:99,Denschlag:00} and bright solitons
   \citep{Khaykovich:02}.
The last result could also be obtained by controlling the dispersion
   of matter waves using a periodic potential \citep{Eiermann:03}.

\subsection{Standard experimental procedures}

A dilute gas of particles can be cooled to a temperature such that the
   de~Broglie wavelength associated to each particle becomes larger
   than the mean interparticle distance.
At these temperatures the quantum statistics of the particles fully
   dominates the behaviour of the gas.
In a trapped atomic gas of bosons all the particles will tend to
   occupy the trap state with the largest population in a very similar
   way to photons in a laser cavity being pulled to the mode with the
   highest gain.
In thermal equilibrium the state with the highest occupancy is the
   ground state of the trap.
In a gas of interacting bosons, however, the ground state of the system
   will not necessarily be the ground state of the potential holding
   the atoms.
In a dilute ultracold atomic gas of bosons only binary collisions are
   allowed and the system is conveniently described by the
   Gross--Pitaevskii equation~\eqref{eq:GPs} for the atomic density
   amplitude.
In particular, it is possible to change the value of the $s$-wave
   scattering length $a$ by different orders of magnitude
   and even change its sign by tuning a magnetic field thanks to
   Fano-Feshbach resonances \citep{Inguscio:03}.
The parameter $a$ strongly influences the properties of the
   condensate.
Condensates with negative $a$ are unstable but can form
   bright solitons.
On the other hand, the ground state of condensates with repulsive
   interactions ($a>0$) significantly differs from the ground state of
   the confining potential.
Indeed, for most experimentally realized condensates the nonlinearity
   totally dominates dispersion, \emph{i.e} the interaction energy is
   much larger than the kinetic energy.
The condensate wavefunction then takes the so called Thomas-Fermi
   shape which has the same symmetry of the trapping potential and,
   for a condensate in a harmonic trap, is an inverted
   parabola \citep{Dalfovo:99}.

The standard experimental procedure for the creation of a
   Bose-Einstein condensate in a dilute atomic gas starts with laser
   cooling of an atomic vapor in ultra-high vacuum conditions.
This first step, performed in a magneto-optical trap \citep{Raab:87},
   takes the atoms from a phase space density of $10^{-20}$ at room
   temperature to $10^{-5}$ below 100~$\mu$K.
At this phase space, density laser cooling stops essentially because
   of spontaneous light scattering from the atoms
   \citep{Chu:98,CohenTannoudji:98,Phillips:98}.
For this reason, the following cooling step has to be performed in a
   non-dissipative trap created either by a magnetic field or by a
   very far-off resonance laser beam.
In non-dissipative traps, cooling is achieved by removing the high
   energy tail of the atomic distribution and by letting the atoms
   thermalize via binary collisions.
Removal of atoms is realized either by reducing the trap depth in
   optical traps or by RF-induced transitions to untrapped Zeeman
   sublevels in magnetic traps.

Let us concentrate on magnetic traps.
The interaction energy of an atom in a magnetic field is $E=-
   \bmu\cdot\bB$, where $\bmu$ is the atomic dipole
   moment.
Of the atomic ground sublevels, only low-field seeking states can be
   trapped in a magnetic field minimum.
If the atomic motion is not very fast, the atomic dipole adiabatically
   follows the magnetic field.
Therefore the energy is proportional to the modulus of the field.
However, if the field variation is very rapid, adiabaticity cannot be
   maintained and atoms are lost from the trap.
To avoid these so called \emph{Majorana spin-flips,} it is necessary
   to avoid a zero of the magnetic field in the trap.
Conventional magnetic traps are realized with a few centimeter size
   coils carrying more than 100~Ampere of current.
In this way traps with $10-100$~Hz oscillation frequencies are
   realized.

The cooling procedure, known as \emph{evaporative cooling,} strongly
   relies on atomic collisions and can be very different for different
   atomic species.
In particular, since the most probable collisions at temperatures
   below 1~mK are spherically symmetric binary collisions,
   due to the Pauli principle \citep{DeMarco:98}, this method cannot
   work for spin-polarized fermions as those trapped in a magnetic
   trap.
Indeed, fermions were cooled to degeneracy either using spin mixtures
   \citep{DeMarco:99} or via collisions with a bosonic atomic species
   \citep{Truscott:01,Schreck:01,Hadzibabic:02,Roati:02}.

\subsection{Optical dipole potentials}

Matter wave beams can be manipulated in very much the same way as
   optical beams.
However, the role of matter and electromagnetic fields is totally
   reversed in the field of atom-optics.
Indeed, condensates can be manipulated with conservative potentials
   created either by far-off resonance laser beams or by magnetic
   fields.
In the first case the potential is obtained via the interaction of the
   induced atomic dipole with the electric field of the laser.
This dipole potential is dependent on the laser intensity and detuning
   and for a two level atom in interaction with a far-off resonance
   beam can be written as
\begin{equation}
V(\br)= \frac{3 \pi c^2}{2
\omega_0^3}\frac{\Gamma}{\Delta}I(\br) ,
\label{dipole}
\end{equation}
where $\omega_0$ is the atomic resonance frequency, $\Gamma$ the
   natural linewidth of the atomic transition, $\Delta$ the laser
   detuning from resonance, and $I(\br)$ the intensity of the
   laserbeam.
From Eq.~\eqref{dipole} we note that when the laser detuning is
   negative, atoms are pulled towards the region with the highest
   laser intensity.
On the other hand, when the detuning is positive, atoms are expelled
   from high intensity regions.
This can be used to create very different kind of potentials with a
   single tunable laser beam.
For example, it is possible to create atomic waveguides either by
   using collimated gaussian laser beams with negative detuning, or
   collimated hollow laser beams with positive detuning
   \citep{Bongs:01}.

\begin{figure}[t]
\centering
\includegraphics[width=0.8\columnwidth]{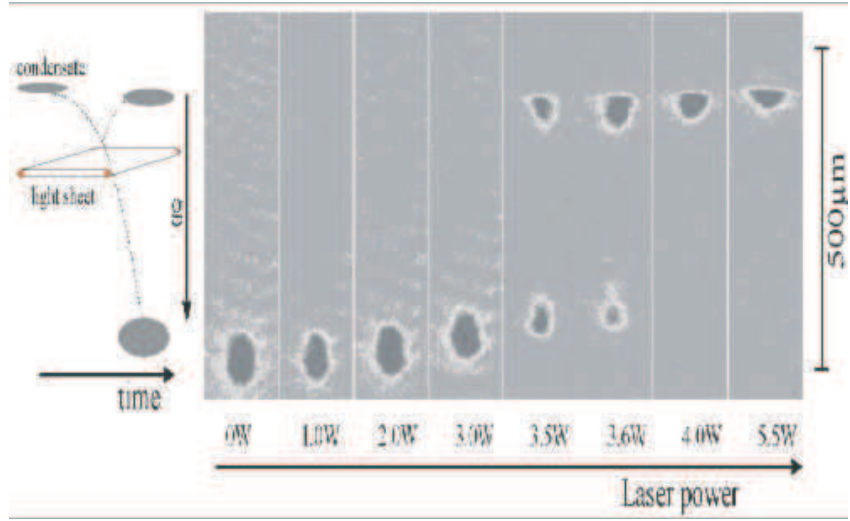}
\caption{Scheme of the IQO experiment.
The condensate falls under the effect of gravity and bounces off the
   potential generated by the laser light.}
\label{fig:IQO}
\end{figure}

With a sheet of light created by rapidly moving a collimated laser
   beam with an acousto-optic modulator, the group of IQO--Hannover
   has been able, by simply varying the laser beam intensity, to
   create an atom mirror, a beam-splitter, or a phase-shifter (see
   Fig.~\ref{fig:IQO}).
The laser detuning for these experiments was positive.
The experiment was performed by creating a condensate via evaporative
   cooling in a magnetic trap \citep{Varenna:98} and then dropping it
   under the effect of gravity onto the sheet of light.
When the laser intensity was very high, the atoms did not acquire
   sufficient kinetic energy during their fall to go over the dipole
   potential and were totally reflected \citep{Bongs:99}.
When the laser intensity was reduced, part of the atoms could go over
   the dipole potential and the system realized a beam-splitter.
By further reducing the laser intensity all the atoms were able to go
   over the potential but were retarded with respect to free fall in
   analogy to an optical phase-shifter \citep{Bongs:01a}.

\subsection{Atom chips and magnetic microtraps}

These examples demonstrate that laser radiation is a very versatile
   tool, since it is possible to create many different potentials
   simply from the interference of laser beams coming from different
   directions as will be discussed below.
On the other hand, it is also possible to manipulate atoms with
   magnetic field gradients.
Magnetic traps are, for neutral atoms, a particularly versatile class
   of these manipulation methods as they can be used for any atomic
   species with a magnetic moment and they can produce conservative
   potentials also for very long times.
Techniques to trap and manipulate atoms with magnetic fields once
   integrated with surface deposition techniques, either lithographic
   or of other kind, realize what is termed an \emph{atom chip}
   \citep{Folman:00}.
\emph{Atom chips} are based on the possibility of creating a 2D-quadrupole
   magnetic field close to a current carrying wire by compensating the
   field generated by the wire $B=\frac{\mu_0}{\pi}\frac{I}{r}$ at
   the height $z_0$ with a constant magnetic field.
If the magnetic moment $\bmu$ of the atom remains aligned with the
   magnetic field the resulting potential for the atoms can be
   approximated as
\begin{equation}
V = \mu\frac{\mu_0 I}{\pi z_0^2}(|z_0-z|+|y|) ,
\end{equation}
where we have assumed the wire to be along the $x$ axis.
With a current of just 0.4~A in the wire and a constant field of 35~G,
   it is possible to create a waveguide with a confining frequency of
   10~kHz for $^{87}$Rb much larger than conventional magnetic traps.

\begin{figure}[t]
\centering
\includegraphics[width=0.8\columnwidth]{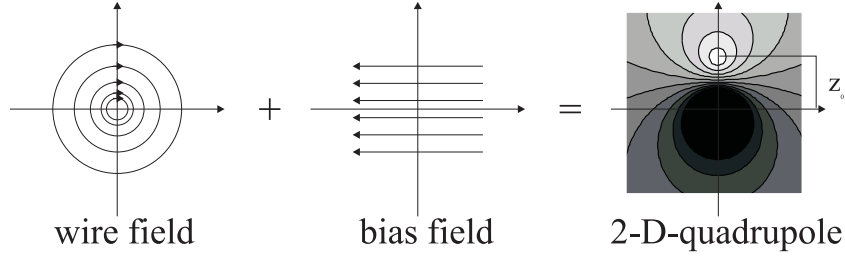}
\caption{Using a bias field to compensate at a point $z_0$ the
   magnetic field generated by a current carrying wire, it is possible
   to create a 2D-quadrupole trap.}
\label{filo}
\end{figure}

\begin{figure}[t]
\centering
\includegraphics[width=0.8\columnwidth]{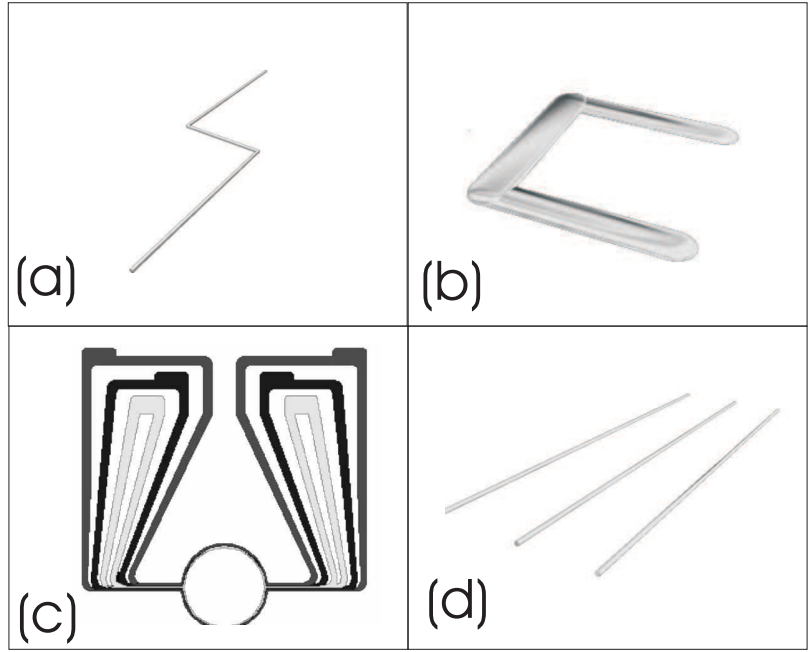}
\caption{Various microtrap configurations: (a) $Z$-Trap; (b) $U$-Trap.
Both the $Z$-Trap and $U$-Trap need an external bias field to achieve a
3D-confining potential for cold atoms; (c) A ring guide with two
diametral potential barriers, working with cold atoms like a SQUID
works with superconductors; (d) A linear guide completely realized
in planar geometry, without any bias field.}
\label{configurazioni}
\end{figure}

The wire guide represents the building brick for magnetic microtraps.
Indeed, by bending the wire in a $U$ shape as illustrated in
   Fig.~\ref{configurazioni}~(b), it is possible to create a 3D
   quadrupole.
As already pointed out, this trap configuration have high atomic
   losses due to Majorana spin-flips.
In order to avoid these losses, it is better to create a harmonic trap
   with a filed minimum different from zero as can be done by bending
   the wire in a $Z$ shape [Fig.~\ref{configurazioni}~(a)].
The bias field can also be realized in a planar configuration using
   three parallel wires with opposite currents as in
   Fig.~\ref{configurazioni}~(d).
This configuration can then be used to create more complicated
   structures as in Fig.~\ref{configurazioni}~(c), where a possible
   analogue of a SQUID is illustrated.

These devices represent one of the most promising schemes for coherent
   atom optics and may be the basis for a totally new class of
   integrated sensors and quantum logic instruments.
Indeed, thanks to component miniaturization, in \emph{atom chips} it
   is possible to reach huge field gradients (above 1~Tesla/cm) with
   1~Ampere currents, of the same order of magnitude as those used in
   electronic circuits and a few orders of magnitude lower than those
   employed in conventional apparatuses.
Furthermore, the substrates employed are compatible with the
   ultra-high vacuum technology needed for atom cooling.
Nowadays the microstructures used experimentally are rather simple and
   the conductors sizes are of the order of 10~$\mu$m with a total
   chip size of a few cm$^2$.
However, they are capable of substituting experimental systems
   currently spread in about 1~m$^3$.
Integrating many elements to control atoms onto a single device, an
   \emph{atom chip}, will make atom physics experiments much more
   robust and simple.
This may allow much more complicated tasks in atom manipulation to be
   performed in a way similar to how integration of electronic
   elements has allowed the development of new powerful electronic
   devices.
Ideally, one desires routine single atom state selective loading,
   preparation and manipulation.
All will be achieved with minimal heat load and power consumption.
Potentials with sizes smaller than the particle de~Broglie wave-length
   will allow tight traps with large energy level spacing.
The large level spacing reduces the probability of the environmental
   noise to induce unwanted excitations.
Consequently, coherent manipulation will be more stable.
Using well developed nanofabrication technology from microelectronics
   to build the atom optics will allow integration of many atom
   optical devices into complex quantum networks combining the best of
   two worlds: the ability to use cold atoms, a well controllable
   quantum system, and the immense technological capabilities of
   nanofabrication and microelectronics to manipulate the atoms.
Experiments based on \emph{atom chips} have allowed the demonstration
   of trapped atom interferometers \citep{Haensel:01a}, of
   \emph{atomic conveyor belts} \citep{Haensel:01}, and of atomic
   waveguides \citep{Folman:00}.
Currently one of the main concerns in \emph{atom chips} is the
   possibility of maintaining a coherent sample close to a surface.
After the achievement of BEC with these devices
   \citep{Haensel:01b,Ott:01}, it was discovered a fragmentation of
   the atomic cloud when brought close to the surface
   \citep{Fortagh:02}.
This fragmentation was very recently attributed to the thickness
   fluctuations of the conductors \citep{Esteve:04}, which causes the
   current to deviate from a straight line producing unwanted axial
   magnetic fields.
Other experiments are concentrating on surface induced losses
   \citep{Jones:03,Harber:03,Lin:04}, however it was possible to
   realize an atomic clock starting from a BEC in an atom chip
   \citep{Treutlein:04}.

\subsection{BEC in periodic potentials}

We now describe some experiments performed on BECs in periodic
   potentials, that well demonstrate the flexibility of optical
   potentials in the manipulation of condensates.
In their experiment, \citet{Fort:00} load cold $^{87}$Rb atoms in the $|F=1 \,
   m_F=-1\rangle$ state from a double magneto-optical trap system into
   a Ioffe type harmonic magnetic trap.
The trap is cylindrically symmetric with an axial frequency of
   $\omega_x/2\pi = 9$~Hz and a radial frequency of
   $\omega_{\perp}/2\pi = 92$~Hz.
\citet{Fort:00} then cool the atoms in this trap via rf-forced
   evaporation until they
   reach a temperature below $300$~nK, just above the critical
   temperature for condensation, which is around $150$~nK dependent on
   the number of atoms we load in the trap.
At this stage of evaporation they suddenly switch on an optical standing
   wave formed by retroreflecting light from a laser blue detuned of
   $\sim 3$~nm with respect to the $D_1$ transition at $\lambda=795$~nm.
The laser beam is aligned horizontally along the axis of the magnetic
   trap and, since the laser beam waist is much larger than the atomic
   cloud transverse size and does not produce any appreciable radial
   force, forms an array of disk shaped traps together with the
   magnetic potential.

The potential is therefore
\begin{equation}
V=\frac{1}{2} m (\omega_x^2 x^2+\omega_{\perp}^2 r_{\perp}^2)+sE_R
\cos^2(2\pi x/\lambda) .
\label{perpot}
\end{equation}
The optical potential is given in units $s$ of the energy $E_R =
   h^2/2m\lambda^2$ gained by an atom (of mass $m$) absorbing one
   lattice photon corresponding in rubidium to a temperature of $\sim
   170$~nK.
In these experiments the optical potential could be varied up to $s=15$.
When $s \gg 1$ the atoms are confined in an array of classically
   independent traps since the optical potential barriers are much
   higher than the thermal energy of the atoms.
Indeed it is possible to ``freeze'' the degree of freedom associated
   with the motion along the axis and study condensation in this quasi
   2D system \citep{Burger:02}.
After switching on the laser light we continue the evaporation ramp
   until the desired temperature is reached.
This ensures that the atoms reach the equilibrium state in the
   combined trap.
When we evaporate to well below the critical temperature, so that no
   thermal fraction is experimentally visible, we typically obtain
   $\sim 200$ condensates separated by a distance of $\lambda/2$, each
   containing $\sim 1000$ atoms.
Due to the blue detuning of the laser beam the atoms are trapped in
   the nodes of the standing wave reducing light scattering below
   $\sim 1$~Hz.

When the height of the optical barriers is much larger than the
   condensates chemical potential we are justified in describing the
   condensate as a sum of wave-functions localized in each potential well:
\begin{equation}
\Psi_0(\br) = \sum_{k=0, \pm 1 \ldots \pm k_M} \exp[-(x-
k\lambda/2)^2/2\sigma^2+i\phi_k] {\sqrt{2} \over g} \mu_k\left(1-
{r^2_{\perp} \over (R_{\perp})^2_k}\right) ,
\label{psi0G}
\end{equation}
where $(R_{\perp})_k = \sqrt{2\mu_k/m\omega^2_{\perp}}$ is the radial
   size of the $k$-th condensate, $g$ depends on the scattering length
   $a$ through Eq.~\eqref{eq:scatt}, while $\mu_k =
   \frac{1}{2} m \omega_x^2d^2 \left(k^2_M-k^2\right)$ plays the role of an
   effective $k$-dependent chemical potential.
The value of $k_M$ is fixed by the normalization condition $N= \sum
   N_k$ to
$$k^2_M = \frac{2 \hbar \overline{\omega} }{ m\omega^2_xd^2 } \left(
   \frac{15}{8\sqrt\pi}  N \frac{a}{a_{ho}}
   \frac{d}{\sigma}\right)^{\frac{2}{5}} ,$$
with
   $\overline{\omega}=(\omega_x\omega_{\perp}^2)^{1/3}$ the
   geometrical average of the magnetic frequencies, $a_{ho} =
   \sqrt{\hbar/m\overline{\omega}}$ is the corresponding oscillator
   length.
From the above equations one also obtains the result $N_k =
   N_0(1-k^2/k^2_M)^2$ with $N_0 = \frac{15}{16} N/k_M$.
Equation~\eqref{psi0G} generalizes the well known Thomas-Fermi results
   holding for magnetically trapped condensates \citep{Dalfovo:99} to
   include the effects of the optical lattice.
This generalization is justified by the fact that the optical
   confinement along the optical lattice is much stronger than the
   magnetic potential therefore it is more suitable to use a harmonic
   approximation for the wave-function along the $x$-direction and a
   Thomas-Fermi approximation in the radial direction \citep{Pedri:01}.

\begin{figure}[t]
\begin{center}
\includegraphics[width=6cm]{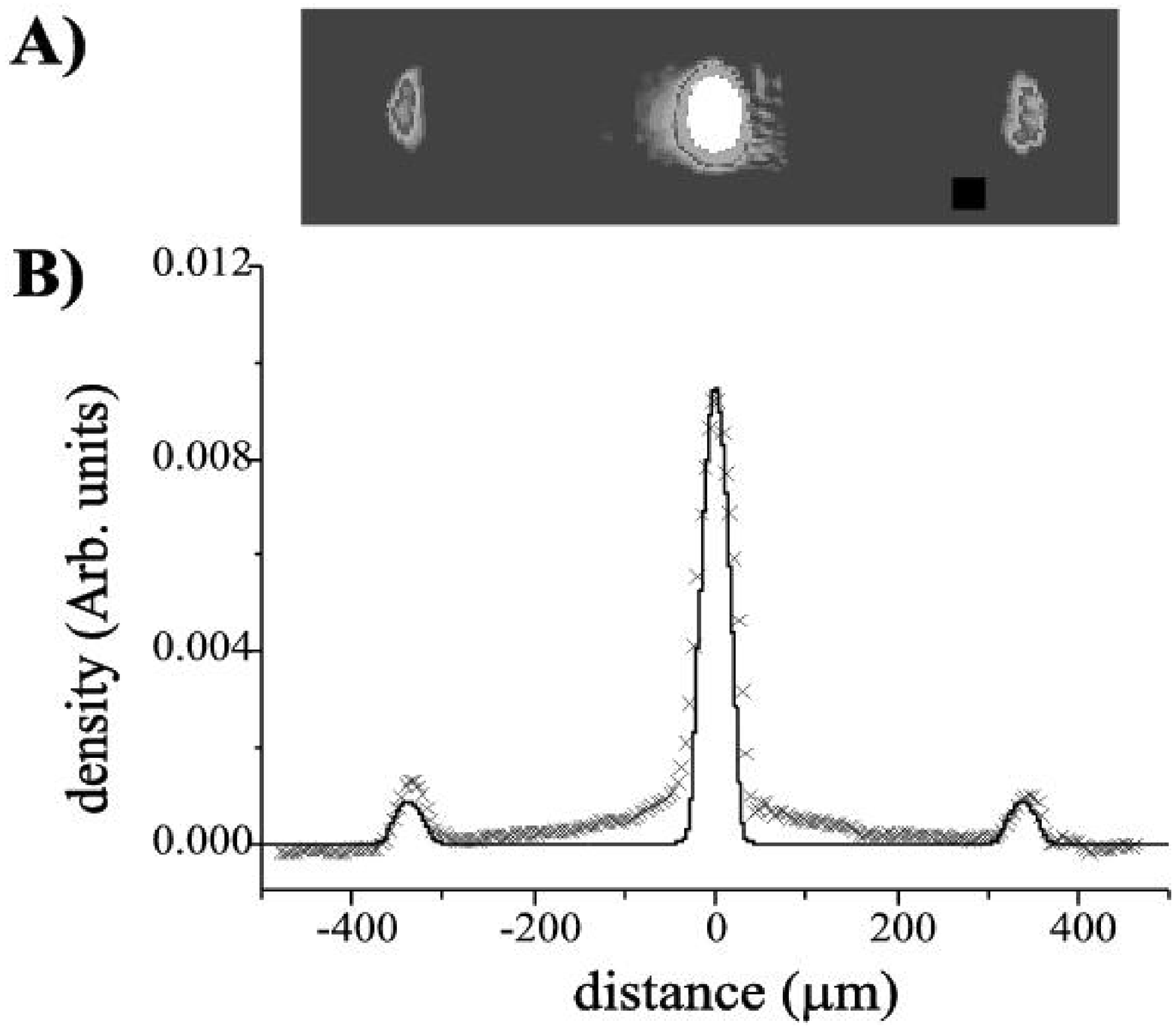}
\includegraphics[width=6cm]{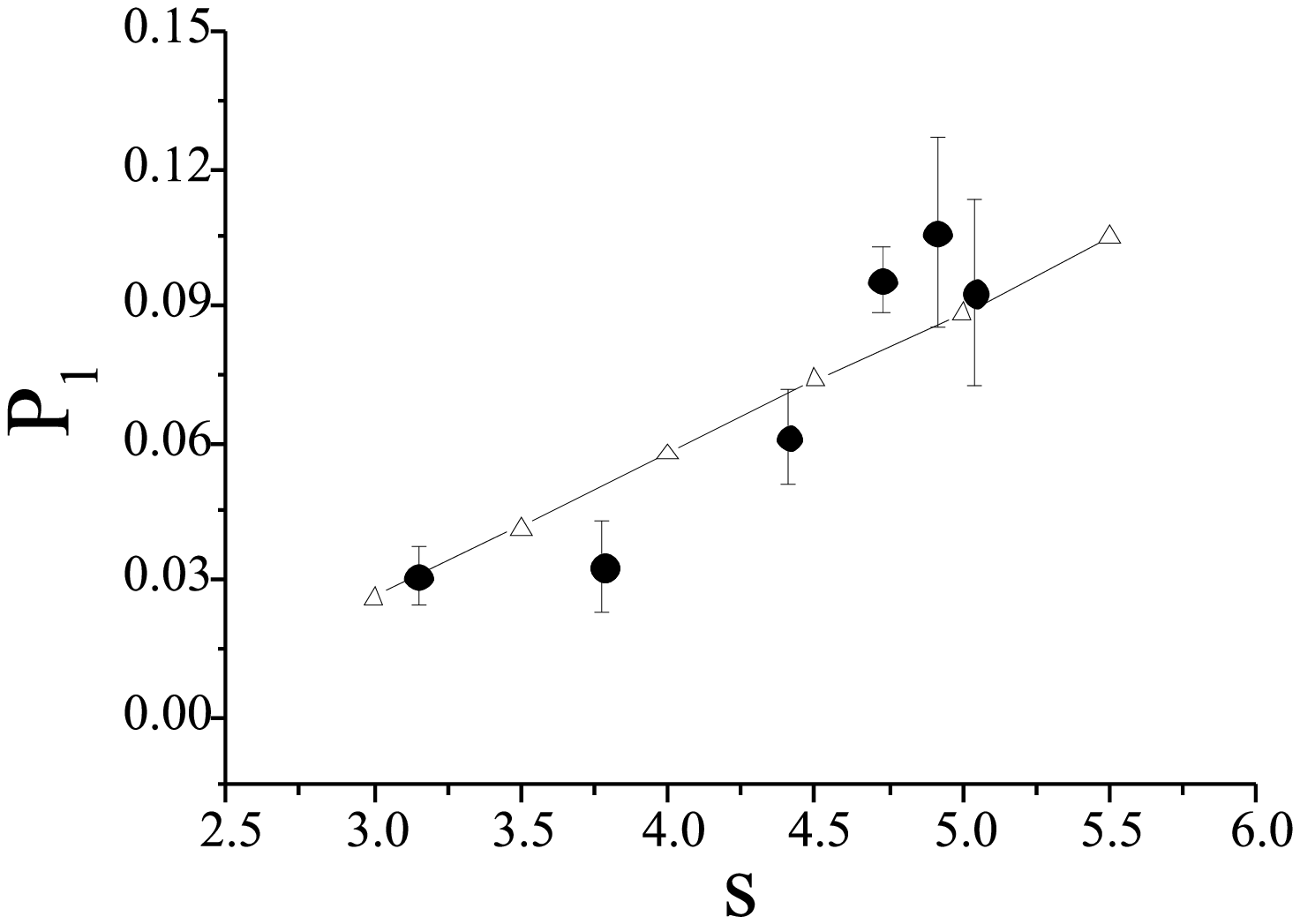}
\end{center}
\caption{\underline{\sl Left:} (A) Absorption image of the density
   distribution of
the expanded array of condensates. (B)  Experimental density
profile (crosses) obtained from the absorption image (A)
integrated along the vertical direction. The wings of the central
peak result from a small thermal component. The continuous line
corresponds to the calculated density profile for the expanded
array of condensates for the experimental parameters ($s=5$ and
$t_{\mathrm{exp}}=29.5$~ms). \underline{\sl Right:} Experimental (circles) and
   theoretical
(triangles) values of the relative population of the $n=1$ peak
with respect to the $n=0$ central one as a function of the
intensity factor $s$ of the optical potential $V_{\mathrm{opt}}$.}
\label{foto}
\end{figure}

When the atoms are released from the combined trap they spread out and
   overlap producing an interferogram which will depend on the
   relative phases $\phi_k$ of the individual condensates.
In Fig.~\ref{foto}~(A), we show a typical image of the cloud taken
   after an expansion time $t_{\mathrm{exp}}=29.5$~ms, corresponding to a total
   number of atoms $N \simeq 20000$ and to an optical potential
   $s=5$.
The image shows a clear structure with three interference peaks
   separated by $2h/m\lambda t_{\mathrm{exp}}$, \emph{i.e.} by the distance
   corresponding to the reciprocal of the lattice constant.
We remark that, differently from the case of two separated
   condensates, interference fringes appear only if the initial
   configuration is mutually coherent.
In other words, since one single interference experiment with an array
   of condensates is equivalent to averaging a series of interference
   experiments with two condensates, an interference pattern will
   appear only in presence of a fixed relative phase between
   condensates belonging to consecutive wells.
What is locking the phase difference across the array of BEC is
   tunnelling through the optical barriers, in a classical picture no
   interference peaks would arise.
The width of the central peak ($n=0$) of the interferogram is of the
   order $\Delta p_x \sim \hbar/mR_x t_{\mathrm{exp}}$ where $R_x \sim k_Md$ is
   half of the length of the whole sample in the $x$-direction.
The occurrence of these peaks is the analogue of multiple order
   interference fringes in light diffraction.

The relative population of the $n \neq 0$ peaks with respect to the
   central one ($n=0$) obeys the simple law
\begin{equation}
P_n=\exp ( -16 \pi^2 n^2 \sigma^2/\lambda^2 ) ,
\label{pop}
\end{equation}
holding also in the presence of a smooth modulation of the atomic
occupation number $N_k$ in each well.
Equation~\eqref{pop} shows that, if $\sigma$ is much smaller than
   $\lambda/2$, the intensity of the lateral peaks will be high, with a
   consequent important layered structure in the density distribution
   of the expanding cloud.
The value of $\sigma$, which characterizes the width of the
   condensates in each well, is determined, in first approximation, by
   the optical confinement.
By using a numerical minimization of the energy we can determine the
   relative population $P_{n}$ of the $n=1$ peak as a function of the
   intensity factor $s$.
This is shown in Fig.~\ref{foto} (right) together with the
   experimental results.
The good comparison between experiment and theory reveals that the
   main features of the observed interference patterns are well
   described by this model.

\subsection{Josephson Junction array with BECs}

In the preceding subsection we have shown that the BECs produced in
   the combined trap are phase-locked by the tunnelling of atoms
   through the optical barriers.
The system indeed realizes a one-dimensional array of Josephson
   Junctions (JJs), as we wish to demonstrate in this section
   \citep{Cataliotti:01}.

A Josephson junction is a simple device made of two coupled
   macroscopic quantum fluids \citep{Barone:82,Barone:00}.
If the coupling is weak enough, an atomic mass current $I$ flows
   across the two systems, driven by their relative phase $\Delta
   \phi$ with a limiting current $I_c$, the ``Josephson critical
   current'', namely the maximal current allowed to flow through the
   junction.
The relative phase dynamics, on the other hand, is sensitive to the
   external and internal forces driving the system being driven by the
   chemical potential difference between the two quantum fluids
   \citep{Tilley:90}.
The arrays of JJs are made of several simple junctions connected in
   various geometrical configurations.
In the last decade, such systems have attracted much interest, due to
   their potential for studying quantum phase transitions in systems
   where the external parameters can be readily tuned
   \citep{Fazio:01}.
Recently, the creation of simple quantum-logic units and more complex
   quantum computer schemes have also been discussed
   \citep{Makhlin:01}.

The condensates in two neighbouring sites of the array have a
   significant interaction via the tunnelling through the barrier, we
   can therefore rewrite the time-dependent Gross-Pitaevskii
   equation~\eqref{eq:GPt}, normally used
   to describe weakly interacting condensates, as a discrete
   non-linear Schr\"odinger (DNLS) equation in a parabolic potential
   \citep{Trombettoni:01}.
We remark that the model used here is one dimensional, as we are now
   only concerned with the motion along the array.
We write the condensate order parameter as $\Psi(x,t)=\sum_j
   \sqrt{N_j(t)} \, e^{i \phi_j (t)} \Phi_j(x)$ and obtain
\begin{equation}
 i  \hbar \frac{\partial \psi_n}{\partial t} = - K
(\psi_{n-1}+\psi_{n+1}) + (\epsilon_n+ \Lambda \mid \psi_n \mid
^2)\psi_n ,
\label{dnls}
\end{equation}
where  $\epsilon_n= \Omega n^2$, $\Omega=\frac{1}{2} m \omega_x^2
   \left(\frac{\lambda}{2} \right)^2$, $\Lambda = g_0 N_T\int dx \,
   \Phi_j^4$.
The tunnelling rate is $K \simeq - \int dx \, \left[ \frac{\hbar^2}{2m}
   \nabla \Phi_j \cdot \nabla \Phi_{j+1} + \Phi_j V \Phi_{j+1} \right]
   $.
We observe that the wavefunctions $\Phi_j$, as well as the tunnelling
   rate $K$, depend on the height of the energy barrier.

In the ground state configuration the Bose-Einstein condensates are
   distributed among the sites at the bottom of the parabolic trap.
If we suddenly displace the magnetic trap along the lattice axis by a
   small distance $\sim 30$~$\mu$m (the dimension of the array is
   $\sim100$~$\mu$m), the cloud will be out of equilibrium and will
   start to move.
As the potential energy that we give to the cloud is still smaller
   than the inter-well barrier, each condensate can move along the
   lattice only by tunnelling through the barriers.
A collective motion can only be established at the price of a well
   definite phase coherence among the condensates.
In other words, the relative phases among all adjacent sites should
   remain locked together in order to preserve the ordering of the
   collective motion.
The locking of the relative phases will again show up in the expanded
   cloud interferogram.

\begin{figure}[t]
\begin{center}
\begin{minipage}[c]{0.35\columnwidth}
\includegraphics[width=\columnwidth]{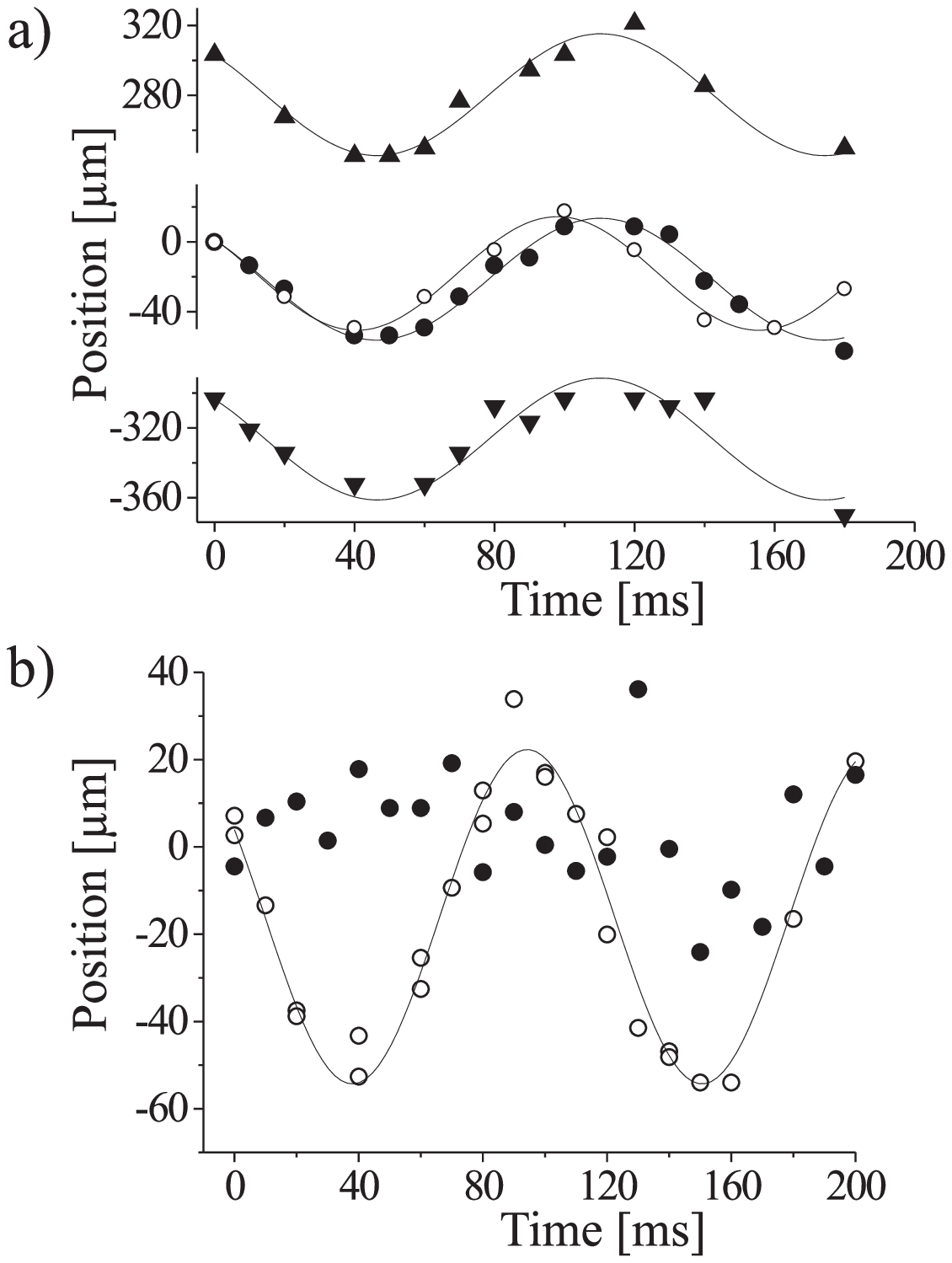}
\end{minipage}
\begin{minipage}[c]{0.35\columnwidth}
\includegraphics[width=\columnwidth]{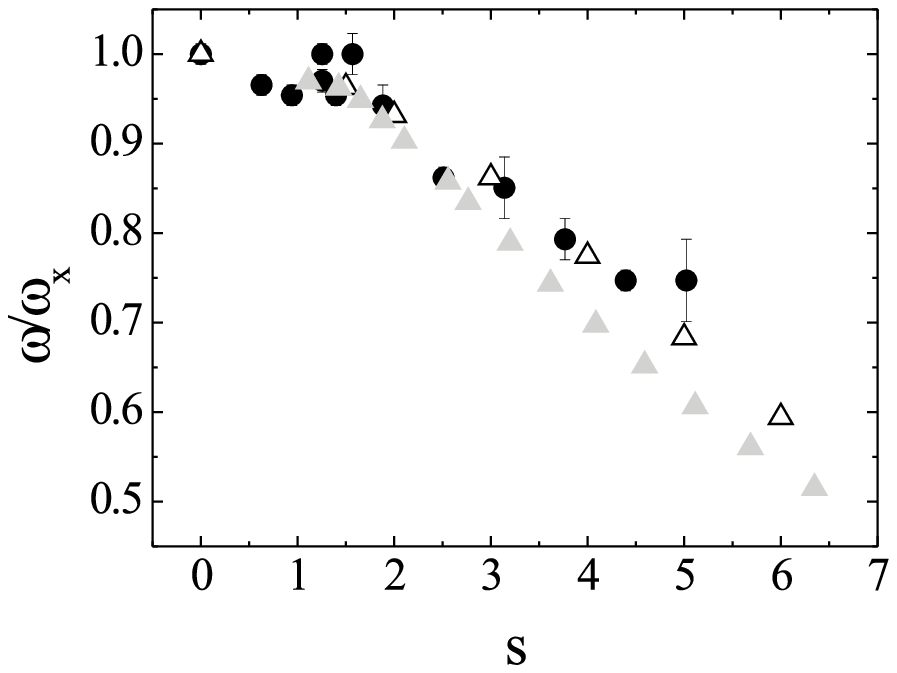}
\end{minipage}
\end{center}
\caption{%
\small
\underline{\sl Left:}
(A) Center of mass positions of the three peaks in
the interferogram of the expanded condensate as a function of the
time
  spent in the combined trap after displacement of the magnetic field.
  Up and down triangles correspond to the first order peaks, filled
  circles to the central peak.  Open circles show the center of mass
  position of the BEC in absence of the optical lattice.  The
  continuous lines are the fits to the data.  (B) Center of mass
  positions of the thermal cloud as a function of time spent in
  the displaced magnetic trap with the standing wave turned on (filled
  circles) and off (open circles).
\underline{\sl Right:} The ratio of the frequency of the atomic current in the
array of Josephson junctions to the harmonic trap frequency as a
function of the inter-well potential height.
  Experimental data (circles) are compared to the values
  calculated with Eq.~\eqref{critical-current} (grey triangles) and to a
   numerical simulation of the 1D GPE (open triangles).}
\label{oscilla}
\end{figure}

For not too large displacements, we observe a coherent collective
   oscillation of the condensates, \emph{i.e.} we see the three peaks
   of the interferogram of the expanded condensates oscillating in
   phase thus showing that the quantum mechanical phase is maintained
   over the entire condensate (Fig.~\ref{oscilla}, left).
In the top part of Figure~\ref{oscilla} we show the positions of the
   three peaks as a function of time spent in the combined trap after
   the displacement of the magnetic trap, compared with the motion of
   the condensate in the same displaced magnetic trap but in absence
   of the optical standing-wave.
(We refer to this as to ``harmonic'' oscillation.)
The motion performed by the center of mass of the condensate is an
   undamped oscillation at a substantially lower frequency than in the
   ``harmonic'' case.
We remark that in a thermal cloud, although individual atoms are
   allowed to tunnel through the barriers, no macroscopic phase is
   present and no motion of the center of mass should be observed.
The center of mass positions of the thermal clouds are also reported
   in Fig.~\ref{oscilla} (left) together with the ``harmonic''
   oscillation of the same cloud in absence of the optical potential.
As can be clearly seen, the thermal cloud does not move from its
   original position in presence of the optical lattice.

The current flowing through the junction between two quantum fluids
   has a maximum value, the critical Josephson current $I_c$, which is
   directly proportional to the tunneling rate $K$.
The existence of such a condition essentially limits the maximum
   velocity at which the condensate can flow through the inter-well
   barriers and therefore reduces the frequency of the oscillations.
As a consequence, we expect a dependence of the oscillation frequency
   on the optical potential through the tunneling rate.
If we rewrite the DNLS equation \eqref{dnls} in terms of the
   canonically conjugated variables population/phase and use
   collective coordinates, we arrive to a phase-current relation
\begin{subequations}
\begin{eqnarray}
\hbar \frac{d}{dt}\xi(t) &=& 2K  \sin{\Delta \phi(t)}   \\
\hbar \frac{d}{dt}\Delta \phi(t) &=& -  m \omega_x^2 \left(
\frac{\lambda}{2} \right)^2  \xi(t) ,
\end{eqnarray}
\label{nc-deltaphi}
\end{subequations}
where $\xi(t)$ is the center of mass of the array and $\Delta \phi(t)$
   the relative phase across the junction.
We remark that, in the regimes we are considering, the current-phase
   dynamics does not depend explicitly on the interatomic
   interaction.
However, it is clear that the non-linear interaction is crucial on
   determining the superfluid nature of the coupled condensates, by
   locking the overall phase coherence against perturbations.

From Eqs.~\ref{nc-deltaphi} we can see that the small amplitude
   oscillation frequency $\omega$ of the current $I \equiv N_T
   \frac{d}{dt}\xi$ gives a direct measurement of the critical
   Josephson current $I_c \equiv 2 K N_T/ \hbar$ and, therefore, of
   the atomic tunneling rate of each condensate through the barriers.
The critical current is related to the frequency $\omega$ of the
   atomic oscillations in the lattice and to the frequency $\omega_x$
   of the condensate oscillations in absence of the periodic field by
   the relation
\begin{equation}
\label{critical-current} I_c = \frac{4 \hbar N_T}{m \lambda^2}
\left( \frac{\omega}{\omega_x} \right)^2.
\end{equation}

Figure~\ref{oscilla} (right) shows the experimental values of the
   oscillation frequencies together with the result of a variational
   calculation.
It must be noted that, due to mean field interactions, in our system
   only for potentials higher than $\sim E_R$ a bound state exists in
   the lattice.

\subsection{Expansion inside a moving 1D optical lattice}

We now discuss a different experiment where we study the expansion of
   a condensate inside a moving 1D optical lattice \citep{Fallani:03}.
This experiment allows us to load the condensate with different
   quasi-momenta $q$ in the periodic structure realized by the
   optical lattice.
We will show how it is possible to adopt a completely different
   language, namely that of band structures in solids, to describe
   the condensate motion. This is possible because the momentum
   spread in the condensate is so small that the behavior of the
   entire atomic cloud can be explained as that of a single
   particle.
From solid state physics it is well known that in the presence of an
   infinite periodic potential the energy spectrum of the free
   particle is modified and a band structure arises \citep[see
   \emph{e.g.}][]{JM1}.
In the rest frame of the lattice the eigenenergies of the system are
   $E_n(q)$, where $q$ is the quasi-momentum and $n$ the band
   index.
According to band theory, the velocity in the $n$-th band is $v_n =\hbar^{-1}
\partial E_n/\partial q$ and the effective mass is $m^\ast=
\hbar^2(\partial ^2 E_n/\partial q^2)^{-1}$.
The effective mass can be negative for a range of quasi-momentum and
   this has been recently recognized as a possibility of realizing
   bright solitons in BEC with repulsive
   interactions \citep{Lenz:94,Konotop:02,Hilligsoe:02,Eiermann:03}.

In this experiment we first produce the condensate in a pure harmonic
   trap, then we switch off the magnetic harmonic potential let the
   BEC expand for 1~ms and we switch on a moving periodic potential.
After 1~ms of expansion the density of the condensate decreases enough
   to neglect the non linear term in the Gross-Pitaevskii equation
   describing an interacting BEC.
This means that, as a first approximation, we are allowed to consider
   the BEC as a linear probe of the periodic potential energy
   spectrum.
The moving periodic potential is created by the interference of two
   counterpropagating laser beams with a slightly different frequency
   and blue detuned 0.5~nm from the $D_2$ resonance at 780~nm.
The two beams are obtained by the same laser and are controlled by two
   independent acousto-optic modulators.
The resulting light field is a standing wave moving in the laboratory
   frame with a velocity $v_L=\lambda \Delta \nu /2$, where $\Delta
   \nu$ is the frequency difference between the two laser beams.
In our experiment, we typically vary the optical lattice velocity
   between 0 and $2v_R$ where $v_R=\hbar k_L/m$ is the recoil velocity
   of an atom absorbing one lattice photon and corresponds, in the
   frame of the band theory, to the limit of the Brillouin zone.
We switch on the moving optical lattice adiabatically by ramping the
   intensity of the two laser beams in 2~ms.
This ensures we are loading the condensate in a Bloch state of
   well-defined energy and quasi momentum \citep{Denschlag:02}.
We let the condensate expand in the lattice and after a total
   expansion time of 13~ms we take an absorption image of the cloud
   along the radial horizontal direction looking at the position and
   dimensions of the condensate inside the optical lattice.
From the position after the expansion, we extract the velocity of the
   condensate inside the optical lattice.
In particular, we repeat the experiment for different velocities of
   the lattice and compare the position of the expanded condensate
   inside the lattice with the position of the condensate expanded
   without the optical lattice.
Let us call $\Delta z$ the difference in position along the axial
   direction.
Then, the velocity of the BEC inside the optical lattice is given by
   $v=\Delta z/\Delta t - v_L$, where $\Delta t$ is the expansion time
   inside the lattice.

\begin{figure}[t]
\centering
\includegraphics[width=0.8\columnwidth]{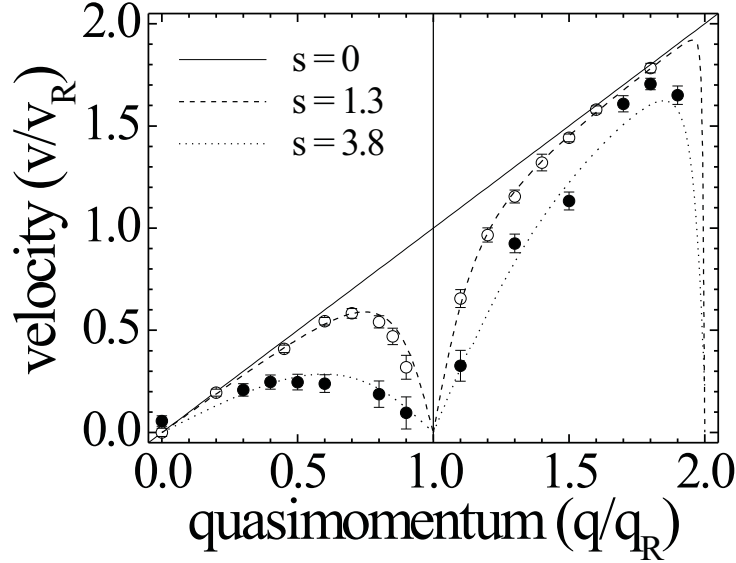}
\caption{Velocity of the condensate inside the optical lattice as
a function of the quasi-momentum $q$ in units of the recoil
momentum $q_R=\hbar k_L$. The open circles corresponds to data
obtained with $V_{\mathrm{opt}}=1.3 E_R$ and the filled circles to data
obtained with $V_{\mathrm{opt}}=3.8 E_R$. The dashed and dotted lines are
the correspondent curves given by the band theory.}
\label{vBloch}
\end{figure}

\begin{figure}[t]
\centering
\includegraphics[width=0.8\columnwidth]{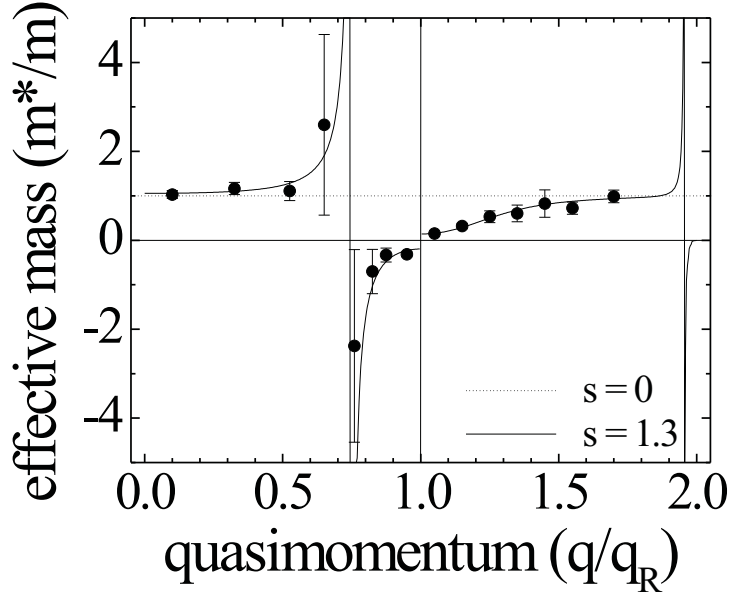}
\caption{Effective mass of a condensate moving in an optical
lattice of $1.3 E_R$ as a function of the quasi-momentum. The data
points correspond to the values extracted from the measured
velocity and the solid line is the corresponding theoretical
prediction of the band theory.}
\label{mBloch}
\end{figure}

In Fig.~\ref{vBloch} we show the results obtained for the velocity of
   the condensate as a function of the quasi-momentum $q$ for two
   different values of the lattice potential depth.
The experimental data points are compared with the theoretical results
   obtained from the band theory and show a very good agreement.
With an adequate sampling of the velocity we can extract the effective
   mass values given by $\partial v/\partial q$.
The results for an optical potential depth of 1.3~$E_R$ are shown in
   Fig.~\ref{mBloch}.
As we increase the lattice velocity (corresponding to increasing the
   quasi-momentum $q$) the effective mass rapidly increases and between
   0.7~$q$ and 0.8~$q$ it first becomes infinite positive and then
   negative.

The consequence of the strong variation of the effective mass is
   expected to consistently modify the expansion of the condensate
   along the axial direction \citep{Massignan:03}.
As a matter of fact, the effective mass enters the diffusive (kinetic)
   term in the Gross-Pitaevskii equation.

\begin{figure}[t]
\centering
\includegraphics[width=0.8\columnwidth]{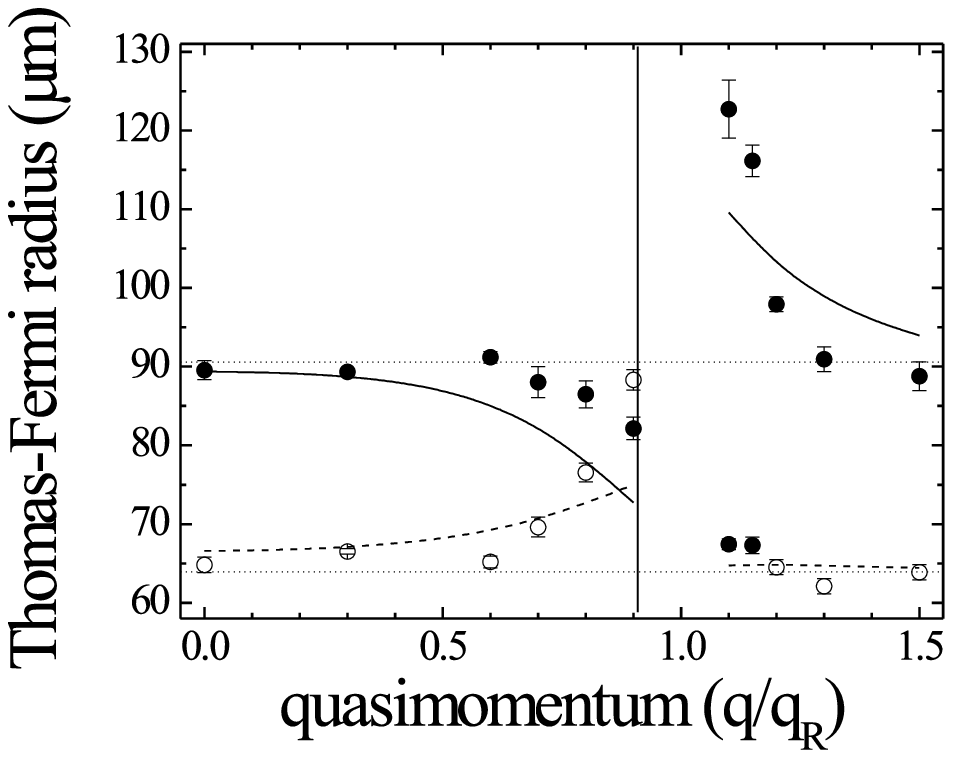}
\caption{Axial and radial dimensions of the condensate after the
expansion in an optical lattice $V_{\mathrm{opt}}=2.9E_R$. The experimental
points (filled and open circles) show the Thomas--Fermi radii of
the cloud extracted from a 2D fit of the density distribution. The
dotted lines show the dimensions of the expanded condensate in the
absence of the optical lattice. The continuous and dashed lines
are theoretical calculations obtained from the 1D effective
model.}
\label{raggiexpa}
\end{figure}

\begin{figure}[t]
\centering
\includegraphics[width=0.8\columnwidth]{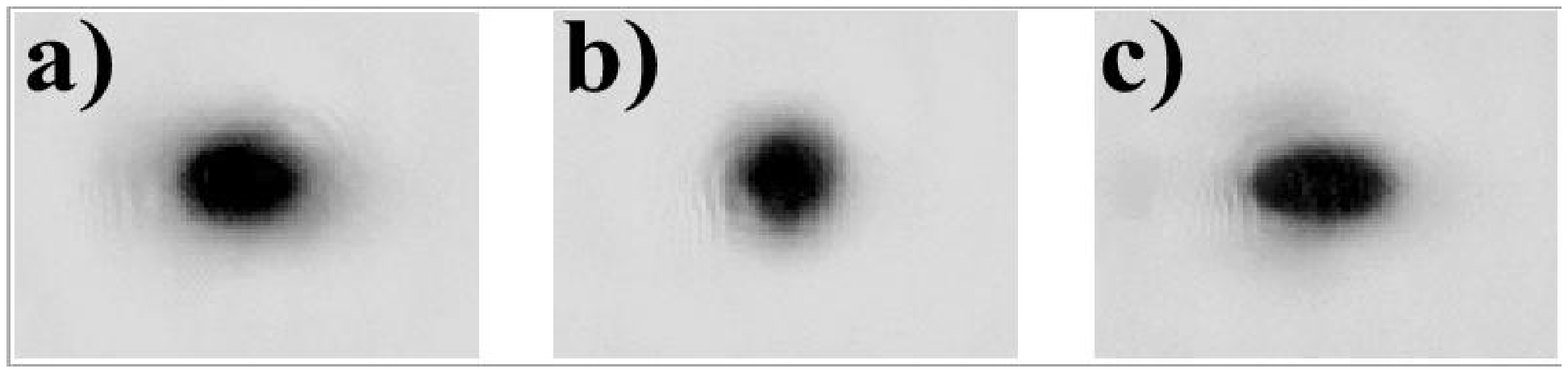}
\caption{Absorption images of the expanded condensate. From left
to right: (a) normal expansion of the condensate without lattice;
(b) axial compression in a lattice of  2.9~$E_R$ and $v_L =
0.9v_R$; (c) enhanced axial expansion in a lattice of 2.9~$E_R$ and
$v_L = 1.1v_R$.}
\label{shapeexpa}
\end{figure}

In Fig.~\ref{raggiexpa} we report the radii of the condensate measured
   as a function of the quasi-momentum after the expansion inside the
   optical lattice compared to numerical predictions based on an
   effective 1D theoretical model \citep{Massignan:03}.
The axial radius (filled circles in Fig.~\ref{raggiexpa}) decreases
   until the quasi-momentum reaches $q_R$.
This is first due to the increase of the effective mass (causing a
   slower expansion) and then by the fact the the effective mass
   becomes negative (causing a contraction of the axial direction
   during the time spent in the optical lattice).

When $q>q_R$, the effective mass becomes positive again but with a
   value smaller then the real mass $m$.
As a consequence the expansion becomes faster in this region of
   quasi-momenta.
In Fig.~\ref{raggiexpa} we also report the measured values of the
   radial dimension of the BEC.
A deviation from the expansion without optical lattice (dashed line)
   is observed also in this direction for $q<q_R$, even if this
   dimension is not directly affected by the presence of the lattice.
This is consistent with the theory (continuous line) and can by
   explained in terms of a coupling between the axial and the radial
   dynamics.

For $q<q_R$, the compression along the lattice direction increases the
   mean-field energy and causes a faster radial expansion.
Instead, when the condensate is loaded with $q>q_R$, the axial
   expansion is enhanced ($0 < m^\ast < m$) and the residual
   mean-field energy is further reduced, causing a suppression of the
   non-linear coupling between the axial and radial dynamics.
This behaviour is evident in the absorption images reported in
   Fig.~\ref{shapeexpa}, where we show the shape of the condensate
   expanded without optical lattice (image \textbf{a)}), and with an
   optical lattice of 2.9~$E_R$ and respectively quasi-momenta $q<q_R$
   (\textbf{b})) and $q>q_R$ (\textbf{c)}).
In the first case, a contraction along the axial direction is
   accompanied by a faster expansion along the radial direction, while
   in the second case the condensate expands faster in the axial
   direction.

\subsection{Experiments beyond the GP equation}

More recently experiments are starting to explore the possibility of
   creating and manipulating pure quantum states of many atoms.
In a beautiful work, \citet{Greiner:02} have has loaded a BEC in a
   three-dimensional optical lattice.
By adiabatically increasing the periodic potential, it was possible to
   enter in a regime where the tunnelling between adjacent wells was
   comparable with the atomic interaction.
In this situation, quantum fluctuations of the atom number in each
   well become relevant and the system can no longer be described by
   the GP equation.
The system undergoes a Mott-insulator quantum phase transition to a
   state where the atom number in each well is fixed
   \citep{Jaksch:98,Sachdev:99}.
In such a situation no interferogram can be recorded when the atoms
   are released from the potential.
In addition, it is not possible to move the atoms through the array
   any longer until the potential energy difference between two
   adjacent wells equals the energy necessary to add an atom to an
   already occupied site.

This experiment, aside from showing the first clear quantum phase
   transition, opened the path for a new set of experiments where it
   was possible to show collapses and revivals of the matter wave
   field, as non-trivial quantum states were formed in the lattice
   \citep{Greiner:02a}.
Later, in the same system, it was possible to create massive
   entanglement between distant atoms in the lattice by the coherent
   manipulation of collisions \citep{Mandel:03} and, more recently, a
   Tonks-Girardeau gas was observed for the first time
   \citep{Paredes:04}.

\section{Limitations of the GP equations from first-principles
   arguments}
\label{sec:limitations}

The purpose of this short section is to expose some limitations of the
   GP equations by appealing to first-principles theory.
The most fundamental approach seems to us to be that due to
   \citet{Leggett:03}.
The arguments outlined below therefore largely follow his article,
   though because of space limitations our discussion will be more
   qualitative than his original study.

His starting point is what he terms the formulation of a
   `pseudo-paradox' in the theory of a dilute Bose gas with repulsive
   interactions.
His `paradox' can be stated as follows.
The usual ground-state energy in the GP approximation set out in
   Section~\ref{sec:GP} above is lower than that in the Bogoliubov
   theory.\footnote{%
Concerning the calculation of the ground-state energy
   of a homogeneous Bose-Einstein condensate beyond Bogoliubov theory,
   we refer the reader to the works of
   \protect\citet{Lieb:98,Weiss:04}, and references given therein.}
Thus, by standard variational arguments, the GP answer should be an
   improved approximation to the Bogoliubov result, which contradicts
   the `established wisdom' concerning this problem.
\citet{Leggett:03} then resolves the `paradox' by a correct
   transcription of the two-body scattering treatment to the many-body
   problem.
He stresses that this resolution has not to do with spurious
   ultraviolet divergences resulting from the replacement of the true
   interatomic potential by a model $\delta$-function pseudopotential.
Instead, \citeauthor{Leggett:03} points out that it comes from an
   infrared divergence which has as a consequence, first of all, that
   the GP approximation actually has no well-defined many-body wave
   function underpinning it.
\citeauthor{Leggett:03} goes on then to show that the `best' attempt
   to construct an approximate variational wave function always
   results in a ground-state energy which either exceeds, or at the
   very best is equal to, that given by the Bogoliubov approximation.

This all prompts us to follow this brief section with an account of a
   proposal by \citet{Angilella:04d} to replace the static GP
   differential Eq.~\eqref{eq:GPs} by an integral equation
   (Sec.~\ref{sec:integral}).
The starting point of their derivation is the Bogoliubov-de~Gennes
   theory, and therefore some physical background to their theory will
   first be given in Section~\ref{sec:BdG} immediately below.

\section{Bogoliubov-de~Gennes equations: physical background}
\label{sec:BdG}

A key ingredient of the GP equation is a spatially varying order
   parameter $\Phi(\br)$, describing the inherently inhomogeneous
   density profile of a BEC condensate.
Such an assumption is usually not necessary for a pure, conventional
   (\emph{i.e.,} $s$-wave) superconductor, for which BCS theory  applies
   with a constant order parameter $\Delta$ \citep[see,
   \emph{e.g.,}][]{Schrieffer:64}.
This is not the case for inhomogeneous superconductors, such as alloys
   or metals in the presence of impurities, both diluted and isolated,
   where a position
   dependent order parameter $\Delta(\br)$ is required.
Such a generalization is also relevant to describe boundary effects in
   superconductors, or the effect of a non-uniform magnetic field,
   as is given by a position dependent vector potential $\bA (\br)$.
The set of coupled equations which relate the
   inhomogeneous pairing potential $\Delta(\br)$ and the
   self-consistent potential $U(\br)$ experienced by single particles
   in the superconductor can be derived within a mean-field approach
   originally due to Bogoliubov \citep{Bogoliubov:59,deGennes:66}.

One usually starts with the effective, mean-field Hamiltonian
\begin{equation}
\hat{H} = \int d\br \hat{\psi}^\dag_\alpha (\br ) \left[ \mathcal{H}_0
   + U(\br) \right] \hat{\psi}_\alpha (\br )
+ \int d\br \left[ \Delta(\br) \hat{\psi}^\dag_\uparrow (\br )
   \hat{\psi}^\dag_\downarrow (\br ) + \mathrm{H.c.} \right]
\label{eq:Heff}
\end{equation}
generalizing the BCS effective Hamiltonian to the inhomogeneous case.
Here, $\hat{\psi}^\dag_\alpha (\br )$ [$\hat{\psi}_\alpha (\br )$]
   is a creation [annihilation] Fermion field operator (with
   understood summations over repeated spin
   indices), $\Delta(\br)$ is
   a spatially varying pairing potential, self-consistently
   taking into account for the electron-electron interaction, which is
   here assumed to be described by a spin independent, strongly local
   potential, as in Eq.~\eqref{eq:gdelta}.
In Eq.~\eqref{eq:Heff}, the kinetic term reads
\begin{equation}
\mathcal{H}_0 = - \frac{\hbar^2}{2m} \left( \nabla - \frac{ie}{\hbar
   c} \bA \right)^2 - \mu ,
\end{equation}
with $\mu$ again the chemical potential and $\bA(\br)$ the electromagnetic
   vector potential, and $U(\br)$ is the self-consistent potential
   experienced by single particles.

One may conveniently expand the field operators over a complete basis
   set as
\begin{equation}
\hat{\psi}_\alpha (\br ) = \sum_\nu \phi_\nu (\br) \hat{c}_{\nu\alpha}
   ,
\end{equation}
where $\hat{c}_{\nu\alpha}$, $\hat{c}^\dag_{\nu\alpha}$ obey the usual
   Fermion anticommutation rules, and $\phi_\nu (\br)$ are
   eigenfunctions of $\mathcal{H}_0$ with
   eigenvalues $\xi_\nu$ (measured from the Fermi level).
Depending on the particular symmetry of the problem under study,
   $\phi_\nu (\br)$ may be plane waves, or more complicated functions
   \citep[see, \emph{e.g.,}][for an instance of axially symmetric
   systems]{Gygi:91}.

Since the effective Hamiltonian, Eq.~\eqref{eq:Heff}, is quadratic in
   the field operators, it can be diagonalized by means of a
   Bogoliubov-Valiatin canonical transformation \citep{Schrieffer:64},
   but now allowing for spatially varying coherent factors $u_\nu (\br)$ and
   $v_\nu (\br)$:
\begin{subequations}
\begin{eqnarray}
\hat{\psi}_\uparrow (\br) &=& \sum_\nu [ u_\nu (\br)
   \hat{\gamma}_{\nu\uparrow} - v_\nu^\ast (\br)
   \hat{\gamma}^\dag_{\nu\downarrow} ] \\
\hat{\psi}_\downarrow (\br) &=& \sum_\nu [ u_\nu (\br)
   \hat{\gamma}_{\nu\downarrow} + v_\nu^\ast (\br)
   \hat{\gamma}^\dag_{\nu\uparrow} ] ,
\end{eqnarray}
\label{eq:BV}
\end{subequations}
with $\hat{\gamma}^\dag_{\nu\alpha}$ [$\hat{\gamma}_{\nu\alpha}$]
   Fermion creation [annihilation] operators for the quasiparticles in
   the superconducting state.

One then requires that Eqs.~\eqref{eq:BV} diagonalize the effective
   Hamiltonian, Eq.~\eqref{eq:Heff}, as
\begin{equation}
\hat{H} = E_0 + \sum_{\nu\alpha} \epsilon_\nu
   \hat{\gamma}^\dag_{\nu\alpha} \hat{\gamma}_{\nu\alpha} ,
\end{equation}
where $E_0$ is the energy of the superconducting ground state.
One then easily finds $[\hat{H},\hat{\gamma}_{\nu\alpha} ] = -
   \epsilon_\nu \hat{\gamma}_{\nu\alpha}$ and
   $[\hat{H},\hat{\gamma}^\dag_{\nu\alpha} ] = \epsilon_\nu
   \hat{\gamma}^\dag_{\nu\alpha}$.
On the other hand, from Eq.~\eqref{eq:Heff}, one has
\begin{subequations}
\begin{eqnarray}
\left[\hat{H},\hat{\psi}_\uparrow (\br)\right] &=& - \mathcal{H}
   \hat{\psi}_\uparrow (\br) - \Delta(\br) \hat{\psi}^\dag_\downarrow
   (\br) \\
\left[\hat{H},\hat{\psi}_\downarrow (\br)\right] &=& - \mathcal{H}
   \hat{\psi}_\downarrow (\br) + \Delta (\br) \hat{\psi}^\dag_\uparrow
   (\br) ,
\end{eqnarray}
\end{subequations}
where $\mathcal{H} = \mathcal{H}_0 + U(\br)$,
and making use of the transformations Eqs.~\eqref{eq:BV} one
   eventually obtains
\begin{subequations}
\begin{eqnarray}
\epsilon_\nu u_\nu (\br) &=& \mathcal{H} u_\nu (\br) +
   \Delta(\br) v_\nu (\br)\\
\epsilon_\nu v_\nu (\br) &=& -\mathcal{H}^\ast v_\nu (\br) +
   \Delta^\ast (\br) u_\nu (\br),
\end{eqnarray}
\label{eq:BdG}
\end{subequations}
which are the static Bogoliubov-de~Gennes (BdG) equations
   \citep{Bogoliubov:59,deGennes:66}.

Eqs.~\eqref{eq:BdG} can be written in compact matrix form as an
   eigenvalue problem for the coherence factors:
\begin{equation}
\begin{pmatrix}
\mathcal{H} & \Delta(\br) \\
\Delta^\ast (\br) & - \mathcal{H}^\ast
\end{pmatrix}
\begin{pmatrix} u_\nu (\br) \\ v_\nu (\br) \end{pmatrix}
= \epsilon_\nu
\begin{pmatrix} u_\nu (\br) \\ v_\nu (\br) \end{pmatrix}.
\end{equation}

Within such a mean-field approach, the pairing potential $\Delta(\br)$
   and the single-particle potential $U(\br)$ are identified with the
   self-consistent statistical averages
\begin{subequations}
\begin{eqnarray}
\Delta(\br ) &=& g \langle \hat{\psi}_\downarrow (br)
   \hat{\psi}_\uparrow (br) \rangle \\
U(\br) &=& g \langle \hat{\psi}^\dag_\uparrow (br)
   \hat{\psi}_\uparrow (br) \rangle,
\end{eqnarray}
\end{subequations}
which, on making use of Eqs.~\eqref{eq:BV}, become:
\begin{subequations}
\begin{eqnarray}
\Delta(\br ) &=& -g \sum_\nu v_\nu^\ast (\br) u_\nu (\br) (1-2f_\nu
   )\\
U(\br) &=& g \sum_\nu \left[ |u_\nu (\br)|^2 f_\nu + |v_\nu (\br) |^2
   (1-f_\nu ) \right],
\end{eqnarray}
\label{eq:selfc}
\end{subequations}
where $f_\nu = [\exp(\epsilon_\nu /\kB T) +1 ]^{-1}$ is the Fermi
   function evaluated at $\epsilon_\nu$.
Thus, superconductivity in an inhomegeneous system is governed by the
   eigenvalue problem given by the BdG Eqs.~\eqref{eq:BdG}, and the two
   self-consistency conditions, Eqs.~\eqref{eq:selfc}.
In analogy with Eq.~\eqref{eq:GPt}, also the BdG equations can be
   generalized to the time dependent case.

One possible variational derivation of the BdG equations rests on the
   generalization of density functional theory (DFT) to
   superconductors \citep{Oliveira:88}.
There, Eqs.~\eqref{eq:BdG} can be identified with the Kohn-Sham equations
   for an appropriate Hohenberg-Kohn functional of both particle density
\begin{equation}
n(\br)= \sum_\sigma \langle \hat{\psi}^\dag_\sigma (\br)
   \hat{\psi}_\sigma (\br) \rangle
\end{equation}
and the superconducting order parameter
\begin{equation}
\chi(\br,\br^\prime ) = \langle \hat{\psi}_\uparrow (\br)
   \hat{\psi}_\downarrow (\br^\prime ) \rangle,
\end{equation}
here generalized to take into account nonlocal effects.
Such an approach has been recently employed to evaluate the
   superconducting condensation energy of the homogeneous electron gas
   with anisotropic pairing potentials, with angular momentum $\ell$
   ranging from 1 to 9 \citep{Wierzbowska:04}.

\section{Integral equation transcending static GP equation taking
   Bogoliubov-de~Gennes theory as starting point}
\label{sec:integral}

In a very stimulating recent contribution, \citet{Pieri:03}
   (referred to as PS below)  have `derived' the
   non-linear GP differential equation for condensed
   bosons by taking
   as their starting point the Bogoliubov-de~Gennes equation for
   superfluid fermions.

The purpose of this section \citep[see also][]{Angilella:04d} is to
   demonstrate that one
   can generalize the zero-temperature differential
   GP equation
   while remaining within the original framework of PS, an
   integral equation formulation then resulting.
The framework of PS is provided by the coupled integral equations involving
   Green functions $\Green_{21}$, $\Green_{11}$ and $\tilde{\Green}_\circ$.
The equations are:
\begin{subequations}
\begin{eqnarray}
\Green_{11} (\br,\br^\prime ;\omega_s ) &=& \tilde{\Green}_\circ
   (\br,\br^\prime ;\omega_s ) + \int \!\! d\br^{\prime\prime} \tilde{\Green}_\circ
   (\br,\br^{\prime\prime} ;\omega_s )
 \Delta(\br^{\prime\prime} )
   \Green_{21} ( \br^{\prime\prime} , \br^\prime ;\omega_s ),\\
\Green_{21} (\br,\br^\prime ;\omega_s ) &=& - \int \!\! d\br^{\prime\prime}
   \tilde{\Green}_\circ
   (\br^{\prime\prime} ,\br ;-\omega_s )
\Delta^\ast (\br^{\prime\prime} )
   \Green_{11} ( \br^{\prime\prime} , \br^\prime ;\omega_s ),
\end{eqnarray}
\label{eq:PS}
\end{subequations}
\!\!\!\!\!
where $\omega_s = (2s+1)\pi/\beta$ ($s$ is an integer) is a fermionic
   Matsubara frequency, $\beta = 1/\kB T$, $\Green_{11}$ is the normal
   and $\Green_{21}$
   is the anomalous single-particle Green function.
The third Green function appearing in Eqs.~\eqref{eq:PS}, namely
   $\tilde{\Green}_\circ$, satisfies the equation
\begin{equation}
[i\omega_s -H(\br) ]\tilde{\Green}_\circ (\br,\br^\prime ; \omega_s ) = \delta
   (\br-\br^\prime ),
\label{eq:H}
\end{equation}
where the single-particle Hamiltonian $H(\br)$ is defined by:
\begin{equation}
H(\br) = - \frac{1}{2m} \nabla^2 + V(\br) -\mu,
\label{eq:H1}
\end{equation}
$\mu$ being the Fermionic chemical potential.
As PS stress, Eqs.~(\ref{eq:PS}), when taken together with the
   self-consistency equation for the gap function:
\begin{equation}
\Delta^\ast (\br) = \frac{V_0}{\beta} \sum_s \Green_{21}
   (\br,\br ; \omega_s ),
\label{eq:gap}
\end{equation}
are entirely equivalent to the Bogoliubov-de~Gennes equations that
   describe the behavior of superfluid fermions in the presence of an
   external potential.
Equations~(\ref{eq:PS}--\ref{eq:gap}) define what we have termed the
   original framework of the PS study.
The constant $V_0 <0$ entering Eq.~(\ref{eq:gap}) arises from the
   contact potential $V_0 \delta(\br -\br^\prime )$ assumed by PS to
   act between fermions with opposite spins.
We also retain here their use of the ratio $\Delta(\br)/\mu$ as an
   expansion parameter which allows the rapid truncation of such
   series, which then leads for strong coupling to an integral
   equation for the gap function
\begin{eqnarray}
-\frac{1}{V_0} \Delta^\ast (\br) &=& \int d\br_1 \, Q(\br,\br_1 )
   \Delta^\ast (\br_1 ) \nonumber\\
&&+\int d\br_1 d\br_2 d\br_3 \, R(\br,\br_1 ,\br_2 ,\br_3 )
   \Delta^\ast (\br_1 ) \Delta(\br_2 ) \Delta^\ast (\br_3 ),
\label{eq:DDD}
\end{eqnarray}
where $R$ is written explicitly in terms of $\tilde{\Green}_\circ
   (\br,\br_1 ;\omega_s )$ in Eq.~(15) of PS.
However, as will emerge below, it is the non-local kernel
   $Q(\br,\br^\prime )$ which is at the heart of the present study.
In terms of the Green function $\tilde{\Green}_\circ$ entering
   Eq.~(\ref{eq:H}), $Q(\br,\br^\prime )$ is given by [PS: Eq.~(14)]:
\begin{equation}
Q(\br,\br^\prime ) = \frac{1}{\beta} \sum_s \tilde{\Green}_\circ
   (\br^\prime ,\br;-\omega_s ) \tilde{\Green}_\circ (\br^\prime ,\br
   ; \omega_s ).
\label{eq:PS14}
\end{equation}
We take the integral equation~(\ref{eq:DDD}) for the gap function as
   the starting point of this section.
For our purposes below, it is then crucial to gain insight into the
   kernel $Q$ in Eq.~(\ref{eq:PS14}), and in particular to carry out
   the summation explicitly over the Matsubara frequencies $\omega_s$.

To gain orientation, let us first perform this summation when the
   external potential $V(\br )$ is set to zero in Eq.~(\ref{eq:H}).
Having achieved this summation, we shall present a general method to
   allow the sum over $\omega_s$ to be achieved for $V(\br)\neq0$,
   using earlier work of \citet{Stoddart:68}.

\begin{figure}[t]
\centering
\includegraphics[height=0.9\columnwidth,angle=-90]{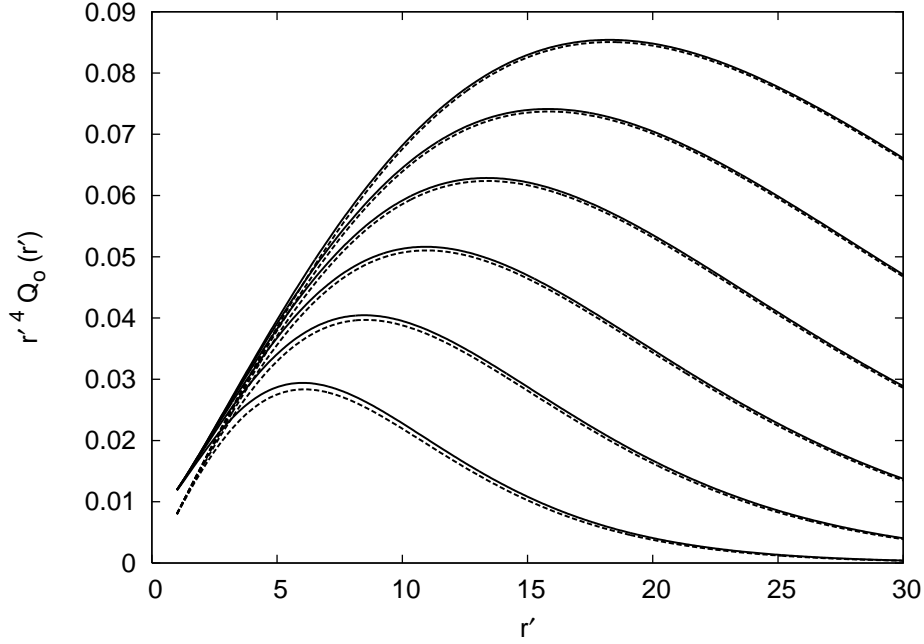}
\caption{Solid lines show $r^{\prime 4} Q_\circ (r^\prime )$, where $Q(r)$
   is defined by Eq.~(\protect\ref{eq:PS14}), as a
   function of $r^\prime = \kF r$, for
   several temperatures, given by $\beta^\prime = \beta\mu=10-30$
   (bottom to top).
Dashed lines are the asymptotic expansion Eq.~(\ref{eq:asymptotic}).
Units are such that $\kF^6/(4\mu) = 1$.
Redrawn from \protect\citet{Angilella:04d}.
}
\label{fig:Q0}
\end{figure}

Returning to the explicit form of $Q(\br,\br_1 )$ given in
   Eq.~(\ref{eq:PS14}) above, it is natural to study first the
   translational invariant, free-electron limit of
   Eq.~(\ref{eq:PS14}), say $Q_\circ (r)$, with
   $r=|\br-\br_1 |$, which is obtained by `switching off' the one-body
   potential $V(\br)$.
This amounts to replacing $\tilde{G}_\circ$ in Eq.~(\ref{eq:PS14}) with
   the free-electron Green function $G_\circ$.
For the Fourier transform of $Q_\circ (r)$, we formally find
\begin{equation}
\hat{Q}_\circ (k) = \int \frac{d \bk^\prime}{(2\pi)^3}
\frac{1-\nF (\xi_{\bk-\bk^\prime} ) - \nF (\xi_{\bk^\prime}
   )}{\xi_{\bk-\bk^\prime} + \xi_{\bk^\prime}} ,
\label{eq:QF0}
\end{equation}
where $\xi_\bk = k^2 /2m - \mu$ and $\nF (\xi)$ is the Fermi-Dirac
   distribution function.
However, it should be noted that, in three dimensions,
   Eq.~(\ref{eq:QF0}) contains a divergent contribution at large
   wave-numbers, which implies a divergent behavior of $Q_\circ (r)$
   at small distances $r$.
Indeed, we find the asymptotic expansion \citep[see also][]{Alexandrov:03}:
\begin{equation}
\frac{4\mu}{\kF^6}
Q_\circ (r)
\sim \frac{1}{4\pi^2} \frac{1}{r^{\prime 2} \beta^\prime } \frac{1}{\sinh a},
\quad \beta^\prime \gg 1,
\label{eq:asymptotic}
\end{equation}
where $r^\prime = \kF r$, $\kF$ is the Fermi wave-number, defined by
   $\mu=\kF^2 /2m$, $\beta^\prime = \beta\mu$, and $a=r^\prime
   \pi/\beta^\prime$.
Fig.~\ref{fig:Q0} shows then our numerical results for $r^{\prime 4} Q_\circ
   (r^\prime )$, as a function of $r^\prime $, for several
   temperatures ($\beta^\prime =10-30$).

Following \citet{Stoddart:68}, the canonical
   density matrix $C(\br,\br^\prime ,\beta)$ is defined by
\begin{equation}
C(\br,\br^\prime ,\beta) = \sum_i \psi_i (\br) \psi_i^\ast (\br^\prime )
   e^{-\beta\epsilon_i} ,
\end{equation}
where $\beta = 1/\kB T$.
Within the perturbative approach of
   \citet{March:60,March:61}, with plane waves as the unperturbed
   solution, the canonical density matrix can then be written to all
   orders in the external potential $V(\br)$ in terms of the
   free-particle canonical density matrix given by
\begin{equation}
C_0 (z,\beta) = (2\pi\beta)^{-3/2} \exp (-z^2 /2\beta ),
\label{eq:C0}
\end{equation}
as
\begin{equation}
C(\br,\br_0 ,\beta) = \int_0^\infty d z\, z\, C_0 (z,\beta)
   f(z,\br,\br_0 ),
\label{eq:C}
\end{equation}
where $f$ satisfies the integral equation \citep{Stoddart:68}:
\begin{equation}
f(z,\br,\br_0 ) = \frac{1}{z} \delta (z-|\br-\br_0 |)
- \int d\br_1 \frac{V(\br_1 )}{2\pi |\br-\br_1 |}
f(z-|\br-\br_1 |,\br_1 ,\br_0 ).
\label{eq:f}
\end{equation}
The desired Green function $\tilde{\Green}_\circ$ is then to be
   obtained from $f$ entering Eqs.~(\ref{eq:C}) and (\ref{eq:f}) as
   \citep{Stoddart:68}
\begin{equation}
\tilde{\Green}_\circ (\br,\br_1 ; k ) = \int_0^\infty d z \, z\,
\bar{\Green}_\circ (z; k) f(z,\br,\br_1 ),
\label{eq:SG}
\end{equation}
where
\begin{equation}
\bar{\Green}_\circ (z;k) = \frac{e^{ikz}}{4\pi z} .
\end{equation}
One may also take advantage of the expression in Eq.~(\ref{eq:SG}) of
   $\tilde{\Green}_\circ$ in terms of $\bar{\Green}_\circ$ to rewrite
   the kernel
   $Q(\br,\br_1 )$ defined by Eq.~(\ref{eq:PS14}) as
\begin{equation}
Q(\br,\br_1 ) = \int_0^\infty dz_1 \, dz_2 \, z_1 \, z_2 \,
f(z_1 ,\br_1 ,\br) f(z_2 ,\br_1 ,\br) Q_\circ (z_1 ,z_2 ),
\label{eq:Qff}
\end{equation}
where the Fourier transform of $Q_\circ (z_1 ,z_2 )$ is given by
\begin{equation}
\hat{Q}_\circ (\bk_1 ,\bk_2 ) = \frac{1-\nF(\xi_{\bk_1} ) - \nF(\xi_{\bk_2}
   )}{\xi_{\bk_1} + \xi_{\bk_2}} .
\end{equation}
Hence, the sum over Matsubara frequencies has still been carried out
   in the presence of an external potential $V(\br)$ entering
   Eq.~(\ref{eq:f}) for the function $f$.

\begin{figure}[t]
\centering
\includegraphics[height=0.9\columnwidth,angle=-90]{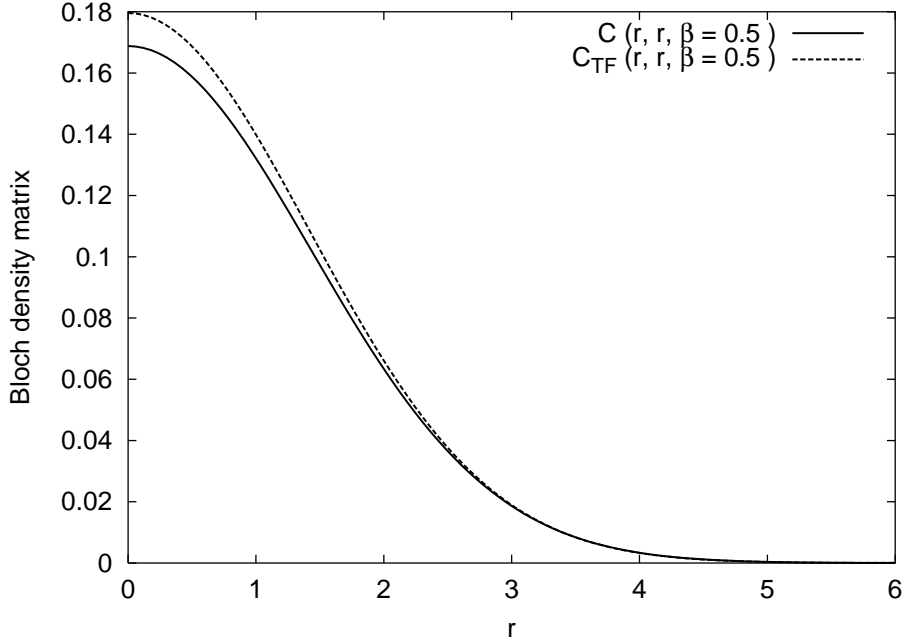}
\caption{Shows diagonal element of the canonical density matrix
   $C(\br,\br,\beta)$, Eq.~(\protect\ref{eq:harmonicC}), and its
   Thomas-Fermi approximation, Eq.~(\protect\ref{eq:harmonicCTF}), as
   a function of $r$, for $\beta=0.5$.
Energies are in units of $\hbar\omega$, while lengths are in units of
   $(\hbar/m\omega)^{1/2}$.
Redrawn from \protect\citet{Angilella:04d}.
}
\label{fig:C}
\end{figure}

Because of current interest in harmonic confinement in magnetic traps
   at low temperatures, let us illustrate the rather formal
   Eqs.~(\ref{eq:C}) and (\ref{eq:f}) when the external potential
   $V(\br)$ has the explicit isotropic harmonic oscillator form in
   three dimensions, namely
\begin{equation}
V(\br) = \frac{1}{2} m\omega^2 r^2 .
\label{eq:harmonic}
\end{equation}
Following the pioneering work of
   \citet{Sondheimer:51} on free electrons in a magnetic field, the
   diagonal element $C(\br,\br,\beta)$ when $V(\br)$ is given by
   Eq.~(\ref{eq:harmonic}) takes the form
   \citep[see \protect\emph{e.g.}][p.~27; see also
   \citeauthor{Howard:03}, \citeyear{Howard:03}]{March:95}
\begin{equation}
C(\br,\br,\beta) = \left( \frac{m}{2\pi\hbar} \right)^{3/2} \left(
   \frac{\omega}{\sinh \hbar\omega\beta} \right)^{3/2}
\exp \left( -\frac{m}{\hbar} \omega r^2 \tanh \frac{1}{2}
   \hbar\omega\beta \right),
\label{eq:harmonicC}
\end{equation}
which is the so-called Slater sum of quantum chemistry (Fig.~\ref{fig:C}).

From Eqs.~(\ref{eq:C0}) and (\ref{eq:C}), performing the substitution
   $t=z^2 /2$, it then follows that
   $f(z,\br,\br_0 )$ can be expressed as the inverse Laplace transform
\begin{equation}
f(z,\br,\br_0 ) = (2\pi)^{3/2} \mathcal{L}^{-1} \left[
s^{-3/2} C(\br,\br_0 ,s^{-1} ) \right]
\label{eq:laplace}
\end{equation}
where $(t,s)$ are conjugate variables with respect to the Laplace
   transform.

Within the Thomas-Fermi (TF) approximation, we take:
\begin{equation}
C_{\mathrm{TF}} (\br,\br,\beta) = \frac{1}{(2\pi\beta)^{3/2}} \exp
   [-\beta V(\br) ],
\label{eq:harmonicCTF}
\end{equation}
which is plotted also in Fig.~\ref{fig:C} for $V(\br)$ given by
   Eq.~(\ref{eq:harmonic}).
For the value of $\beta$ shown, the TF form Eq.~(\ref{eq:harmonicCTF})
   is seen to be a useful approximation to the exact result,
   Eq.~(\ref{eq:harmonicC}).
Inserting Eq.~(\ref{eq:harmonicCTF}) into Eq.~(\ref{eq:laplace}) we
   find
\begin{equation}
f_{\mathrm{TF}} (z,\br,\br) = \frac{\delta(z)}{z} -
   \frac{\sqrt{2V(\br)}}{z} J_1 [\sqrt{2V(\br)} z ],
\label{eq:fTF}
\end{equation}
where $J_1$ denotes the Bessel function of the first kind and order
   one.

\begin{figure}[t]
\centering
\includegraphics[height=0.9\columnwidth,angle=-90]{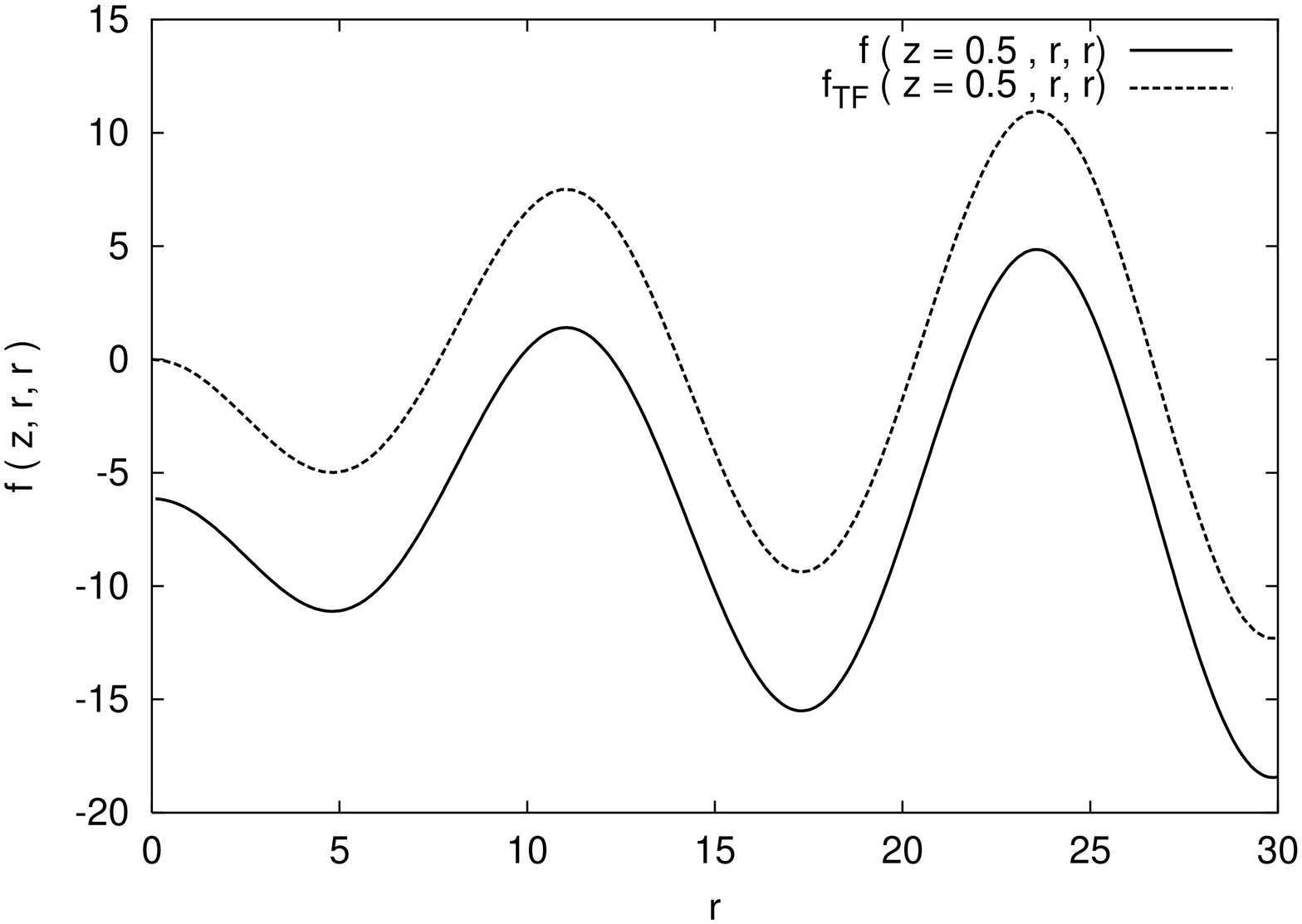}
\caption{Shows diagonal $f(z,\br,\br)$ corresponding to the harmonic
   potential, as given by the inverse Laplace transform,
   Eq.~(\protect\ref{eq:laplace}), as well as the regular part of its
   Thomas-Fermi approximation, Eq.~(\ref{eq:fTF}), as a function of
   $r$, for fixed $z=0.5$.
Units as in Fig.~\protect\ref{fig:C}.
Redrawn from \protect\citet{Angilella:04d}.}
\label{fig:f}
\end{figure}

Fig.~\ref{fig:f} shows $f(z,\br,\br)$ as a function of $r$ for fixed
   $z$, as obtained by numerically performing the inverse Laplace
   transform in Eq.~(\ref{eq:laplace}) for the harmonic potential
   case.
The regular contribution to the analytic result for the Thomas-Fermi
   approximation, Eq.~(\ref{eq:fTF}), is also plotted for comparison.
The similarity in shape between approximate and (numerically) exact
   results for this harmonic confinement model seems to us rather
   remarkable.
After this model test of a TF-like approximation invoked by PS, we
   return to the general case, based on the exact result
   Eq.~(\ref{eq:Qff}) for the kernel $Q(\br,\br_1 )$.

Then, we invert the argument of PS but still use a further essential
   assumption of their study, namely that the condensate wave function
   $\Phi(\br)$ entering the Gross-Pitaevskii equation is related to
   the gap function $\Delta(\br)$ by
\begin{equation}
\Phi(\br) = \left( \frac{m^2 a_{\mathrm{F}}}{8\pi} \right)^{1/2}
   \Delta(\br) \equiv k \Delta(\br).
\label{eq:k}
\end{equation}
Here, in the strong coupling limit, and following PS, $a_{\mathrm{F}}
   \sim (2m|\mu|)^{-1/2}$ represents the characteristic length scale
   for the non-interacting Green function $\Green_\circ$, equal to
   $\tilde{\Green}_\circ$ above when $V(\br)$ is put equal to zero.

Given the validity of this PS assumption, Eq.~(\ref{eq:k}), we then
   rewrite Eq.~(\ref{eq:DDD}) as an equation for $\Phi(\br)$:
\begin{eqnarray}
-\frac{1}{V_0} \Phi^\ast (\br) &=& \int d\br_1 \, Q(\br,\br_1 )
   \Phi^\ast (\br_1 ) \nonumber\\
&&+\frac{1}{k^2}
\int d\br_1 d\br_2 d\br_3 \, R(\br,\br_1 ,\br_2 ,\br_3 )
   \Phi^\ast (\br_1 ) \Phi(\br_2 ) \Phi^\ast (\br_3 ).
\label{eq:PPP}
\end{eqnarray}
This then is the proposed generalization of the Gross-Pitaevskii
   equation, but with $Q(\br,\br_1 )$ to be calculated more
   accurately than by the Thomas-Fermi--like assumption of
   \citet{Pieri:03}, via Eqs.~(\ref{eq:Qff}) and (\ref{eq:f}).

While Eq.~(\ref{eq:PPP}) is a direct consequence of the above
   arguments, it remains an expansion in $\Phi$, in suitable reduced
   form.
Therefore, a first attempt to simplify this Eq.~(\ref{eq:PPP}) is to
   retain the approximation given by the Pieri-Strinati approach in the
   `smallest' term involving $O(\Phi^3 )$ on the right-hand side of
   the basic Eq.~(\ref{eq:PPP}).
Thus one reaches the (still non-local) equation for the condensate
   wave function $\Phi(\br)$:
\begin{equation}
-\frac{1}{V_0} \Phi^\ast (\br) = \int d\br_1 \, Q(\br,\br_1 )
   \Phi (\br_1 )
-\frac{1}{2} ma_{\mathrm{F}}^2 |\Phi(\br)|^2 \Phi(\br).
\label{eq:PPPP}
\end{equation}

For sufficiently small spatial variations in the condensate wave
   function $\Phi(\br)$ in Eq.~(\ref{eq:PPPP}), the basic nonlocality
   can be removed by Taylor expanding $\Phi(\br_1 )$ around the
   position $\br$ in the integral term.
This then characterizes the problem in terms of `partial moments' of
   the kernel $Q(\br,\br_1 )$, namely $\smallint Q(\br,\br_1 )d\br_1$
   and $\smallint Q(\br,\br_1 )|\br-\br_1 |^2 d\br_1$.
Such partial moments then enter the original GP
   equation, as stressed by PS.

In summary, we propose the retention of the non-local kernel
   $Q(\br,\br_1 )$ as in Eq.~(\ref{eq:PPP}) above, since the sum over
   Matsubara frequencies in Eq.~(\ref{eq:PS14}) has been performed in
   Eq.~(\ref{eq:Qff}), which is a central result of the present study.
However, in the terms of $O(\Phi^3 )$ in Eq.~(\ref{eq:PPP}), a
   sensible starting point is to follow the PS approximation displayed
   in Eq.~(\ref{eq:PPPP}).

As to future directions, evaluation of the non-local kernel in
   Eq.~(\ref{eq:Qff}) for other external potentials than the harmonic
   case in Eq.~(\ref{eq:harmonic}) is of obvious interest.
For this latter model, though our Fig.~\ref{fig:f} considers the
   diagonal element of $f(z,\br ,\br_1 )$, the off-diagonal form of
   $C(\br,\br_1 ,\beta)$ is known \citep{Howard:03}, and numerical
   Laplace inversion to obtain $f(z,\br ,\br_1 )$ is entirely feasible.
Then $Q(\br,\br_1 )$ can be obtained, though of course numerically.

The GP equation is valid in the strong-coupling limit of
   superfluidity.
It has to be stressed that in the weak coupling limit one can
   also derive a Ginzburg-Landau equation starting from the
   Bogoliubov-de~Gennes equations.
We note specifically in this context that the derivation of the
   Ginzburg-Landau equation in the weak-coupling limit for the
   harmonic trap was presented by \citet{Baranov:98}.
The results presented in this section
   are also relevant to the weak-coupling limit of
   superfluidity.

Finally, we return to the foundations of the Gross-Pitaevskii equation
   by \citet{Leggett:03}, which we summarized very briefly in
   Section~\ref{sec:limitations} above.
\citeauthor{Leggett:03} concludes that there is no correlated
   many-body wave-function underlying their original equation.
It will be interesting for the future to know whether the non-local
   versions of Eqs.~(\ref{eq:PPP}) and (\ref{eq:PPPP}) proposed here
   are still subject to this limitation.

\section{Summary and suggestions for further study}
\label{sec:summary}

In summary, the GP Eqs.~\eqref{eq:GPs} and \eqref{eq:GPt} have been
   considered in relation to a variety of experiments on inhomogeneous
   condensed Bosons in Sections~\ref{sec:GP} first and, predominantly,
   in Section~\ref{sec:expt}.

Sections~\ref{sec:limitations}--\ref{sec:integral} are then
   theoretical in emphasis.
The short Section~\ref{sec:limitations} summarizes the arguments of
   \citet{Leggett:03}, which reveal that there is no underlying
   many-body wave function for the static GP Eq.~\eqref{eq:GPs}.
Therefore, instead of the GP differential equation,
   Sections~\ref{sec:BdG} and \ref{sec:integral} present an integral
   equation theory \citep{Angilella:04d} for which the starting point
   is the Bogoliubov-de~Gennes theory.

\subsection{Suggestions for further study}

Though a non-local theory given in Section~\ref{sec:integral}, in the
   form of an integral equation, trascends the static GP differential
   Eq.~\eqref{eq:GPs}, no proof has, as yet, been given that this
   integral equation theory has, underlying it, a many-body wave
   function.
It seems that this, therefore, in view of Leggett's criticism of the
   foundations of the static GP Eq.~\eqref{eq:GPs}, is an area of
   considerable interest for further theoretical work.

\citet{ODell:04} have developed, within the Thomas-Fermi regime, the
   hydrodynamics of a trapped dipolar BEC.
As they stress, such a BEC, whose particles interact via dipole-dipole
   coupling, constitute an example of a superfluid with long-range and
   anisotropic interparticle forces.
This contrasts with the customary BECs which are characterized by
   isotropic interactions with range much smaller than the average
   interparticle interaction \citep[see \emph{e.g.}][]{Dalfovo:99}.
While alkali atoms have small dipole-dipole interactions,
   \citeauthor{ODell:04} suggest that chromium is a promising case for
   a dipolar BEC, due to its large magnetic moment of 6~Bohr
   magnetons, and there has been progress in cooling it towards
   degeneracy \citep{Schmidt:03,Hensler:03}.
As \citeauthor{ODell:04} also note, molecules can possess large dipole
   moments.
They suggest that advances in the cooling of polar molecules
   \citep{Weinstein:98,Bethlem:99,Bethlem:00,Bethlem:02},
   photoassociation of ultracold heteronuclear molecules
   \citep{Mancini:04}, and
   molecular BECs \citep{Greiner:03,Jochim:03,Zwierlein:03} may soon
   lead to superfluids where dipolar effects play a major role.

\citet{ODell:04} consider in their Letter a harmonically trapped BEC
   with dipole-dipole interactions as well as short-range $s$-wave
   scattering.
They note that in the Thomas-Fermi limit, where the zero-point kinetic
   energy of the atoms in the trap can be neglected compared with the
   interparticle interaction energy and the trapping potential, the
   collective dynamics of a BEC may be treated by the collisionless
   hydrodynamic theory of Bose superfluids at $T=0$
   \citep{Stringari:96}.
The effect of the dipolar interactions is to introduce non-locality
   into the already nonlinear hydrodynamic equations.

In relation to the focus of the present review on the GP equations, it
   is to be emphasized \citep{ODell:04} that the dominant interactions
   in the ultracold gases available at the time of writing are
   asymptotically of van~der~Waals character, with an $r^{-6}$ decay,
   and are short-ranged, compared with the mean interactomic distance.
As we have seen above, within the mean-field regime of the GP
   equations these interactions are represented by the pseudopotential
   model $g\delta(\br) \equiv (4\pi a_s \hbar^2 /m )\delta(\br)$
   [cf. Eqs.~\eqref{eq:gdelta} and \eqref{eq:scatt}].
By employing a Fano-Feshbach resonance, it is possible to adjust the
   value of $a_s$ to yield repulsive (positive) and attractive
   (negative) values \citep{Inouye:98}.
Appealing to the analogy of nuclear magnetic resonance techniques,
   dipole-dipole interactions can also be controlled by rapidly
   rotating an external field \citep{Giovanazzi:02}.
The interactions vanish when the rotation corresponds to the so-called
   magic angle.

To summarize here the main points made by \citet{ODell:04}, the
   versatility of quantum gases is promising for the study of the role
   of interactions in superfluidity.
They present solutions of the dipolar superfluid hydrodynamic
   equations in a harmonic trap: the condensate density is parabolic
   as in the pure $s$-wave problem but now with modified radii.

Finally, and returning to the Fano-Feshbach resonances already
   referred to \citep{Fano:35,Feshbach:58}, there remains the
   important question for theory of the crossover of Bosonic to
   Fermionic superfluidity \citep[see, \emph{e.g.,}][]{Randeria:95},
   following, for example, the pioneering studies of
   \citet{Nozieres:85} \citep[see also][]{Timmermans:01}.
Note that the cold-atom technology that has resulted in the
   observation of dilute-gas BECs enables the creation and subsequent
   studies of physical properties of novel superfluids having a degree
   of flexibility adding to the already powerful methods of
   traditional low temperature physics.
This therefore seems a further, and very attractive, direction for
   further studies.

\subsection*{Acknowledgements}

The authors would like to thank N.~Andrenacci, M.~Baldo, M.~L.~Chiofalo,
   I.~A.~Howard, M.~Inguscio, E.~Rimini, and M.~P.~Tosi for
   useful discussions on the general area embraced by the present
   review.

\bibliographystyle{philmagb}
\bibliography{a,b,c,d,e,f,g,h,i,j,k,l,m,n,o,p,q,r,s,t,u,v,w,x,y,z,zzproceedings,Angilella}

\begin{thebibliography}{}

\bibitem[Alexandrov, 2003]{Alexandrov:03}
\authorfont{Alexandrov, A.~S.}
\newblock {\em Theory of superconductivity: From Weak to Strong Coupling} (IOP,
  Bristol, 2003).

\bibitem[Anderson and Kasevich, 1998]{Anderson:98e}
\authorfont{Anderson, B.~P. and Kasevich, M.~A.} (1998).
\newblock Science {\bf 281}, 1686.

\bibitem[Andrews {\em et~al.}, 1997]{Andrews:97}
\authorfont{Andrews, M.~R., Townsend, C.~G., Miesner, H.~J., Durfee, D.~S.,
  Kurn, D.~M., and Ketterle, W.} (1997).
\newblock Science {\bf 275}, 637.

\bibitem[Angilella {\em et~al.}, 2004]{Angilella:04d}
\authorfont{Angilella, G. G.~N., March, N.~H., and Pucci, R.} (2004).
\newblock Phys. Rev. A {\bf 69}, 055601.

\bibitem[Baranov and Petrov, 1998]{Baranov:98}
\authorfont{Baranov, M.~A. and Petrov, D.~S.} (1998).
\newblock Phys. Rev. A {\bf 58}, R801.

\bibitem[Barone, 2000]{Barone:00}
\authorfont{Barone, A.} (2000).
\newblock Weakly coupled macroscopic quantum systems: likeness with difference.
\newblock In {\em Quantum Mesoscopic Phenomena and Mesoscopic Devices in
  Microelectronics}, edited by Kulik, I.~O. and Ellialtioglu, R., volume 559 of
  {\em NATO Science Series: C Mathematical and Physical Sciences}, p. 301
  (Kluwer, Dordrecht, 2000).

\bibitem[Barone and Paterno, 1982]{Barone:82}
\authorfont{Barone, A. and Paterno, G.}
\newblock {\em Physics and Applications of the {Josephson} Effect} (J. Wiley \&
  Sons, New York, 1982).

\bibitem[Bethlem {\em et~al.}, 2000]{Bethlem:00}
\authorfont{Bethlem, H.~L., Berden, G., Crompvoets, F. M.~H., Jongma, R.~T.,
  {van Roij}, A. J.~A., and Meijer, G.} (2000).
\newblock Nature (London) {\bf 406}, 491.

\bibitem[Bethlem {\em et~al.}, 1999]{Bethlem:99}
\authorfont{Bethlem, H.~L., Berden, G., and Meijer, G.} (1999).
\newblock Phys. Rev. Lett. {\bf 83}, 1558.

\bibitem[Bethlem {\em et~al.}, 2002]{Bethlem:02}
\authorfont{Bethlem, H.~L., Crompvoets, F. M.~H., Jongma, R.~T., {van de
  Meerakker}, S. Y.~T., and Meijer, G.} (2002).
\newblock Phys. Rev. A {\bf 65}, 053416.

\bibitem[Bloch {\em et~al.}, 1999]{Bloch:99}
\authorfont{Bloch, I., H\"ansch, T.~W., and Esslinger, T.} (1999).
\newblock Phys. Rev. Lett. {\bf 82}, 3008.

\bibitem[Bogoliubov, 1947]{Bogoliubov:47}
\authorfont{Bogoliubov, N.~N.} (1947).
\newblock J. Phys. (Moscow) {\bf 11}, 23.

\bibitem[Bogoliubov {\em et~al.}, 1959]{Bogoliubov:59}
\authorfont{Bogoliubov, N.~N., Tolmachev, V.~V., and Shirkov, D.~V.}
\newblock {\em A New Method of Superconductivity} (Consultants Bureau, New
  York, 1959).

\bibitem[Bongs {\em et~al.}, 1999]{Bongs:99}
\authorfont{Bongs, K., Burger, S., Birkl, G., Sengstock, K., Ertmer, W.,
  Rzazewski, K., Sanpera, A., and Lewenstein, M.} (1999).
\newblock Phys. Rev. Lett. {\bf 83}, 3577.

\bibitem[Bongs {\em et~al.}, 2001a]{Bongs:01a}
\authorfont{Bongs, K., Burger, S., Dettmer, S., Hellweg, D., Arlt, J., Ertmer,
  W., and Sengstock, K.} (2001a).
\newblock Cr. Acad. Scie. IV-Phys. {\bf 2}, 671.

\bibitem[Bongs {\em et~al.}, 2001b]{Bongs:01}
\authorfont{Bongs, K., Burger, S., Dettmer, S., Hellweg, D., Arlt, J., Ertmer,
  W., and Sengstock, K.} (2001b).
\newblock Phys. Rev. A {\bf 63}, 031602(R).

\bibitem[Burger {\em et~al.}, 1999]{Burger:99}
\authorfont{Burger, S., Bongs, K., Dettmer, S., Ertmer, W., Sengstock, K.,
  Sanpera, A., Shlyapnikov, G.~V., and Lewenstein, M.} (1999).
\newblock Phys. Rev. Lett. {\bf 83}, 5198.

\bibitem[Burger {\em et~al.}, 2002]{Burger:02}
\authorfont{Burger, S., Cataliotti, F.~S., Fort, C., Maddaloni, P., Minardi,
  F., and Inguscio, M.} (2002).
\newblock Europhys. Lett. {\bf 57}, 1.

\bibitem[Burger {\em et~al.}, 2001]{Burger:01}
\authorfont{Burger, S., Cataliotti, F.~S., Fort, C., Minardi, F., Inguscio, M.,
  Chiofalo, M.~L., and Tosi, M.~P.} (2001).
\newblock Phys. Rev. Lett. {\bf 86}, 4447.

\bibitem[Castin and Dum, 1999]{Castin:99}
\authorfont{Castin, Y. and Dum, R.} (1999).
\newblock Eur. Phys. J. D {\bf 7}, 399.

\bibitem[Cataliotti {\em et~al.}, 2001]{Cataliotti:01}
\authorfont{Cataliotti, F.~S., Burger, S., Fort, C., Maddaloni, P., Minardi,
  F., Trombettoni, A., Smerzi, A., and Inguscio, M.} (2001).
\newblock Science {\bf 293}, 843.

\bibitem[Cherny and Brand, 2004]{Cherny:04}
\authorfont{Cherny, A.~Y. and Brand, J.} (2004).
\newblock Phys. Rev. A {\bf 70}, 043622.

\bibitem[Chiofalo {\em et~al.}, 1999]{Chiofalo:99}
\authorfont{Chiofalo, M.~L., Succi, S., and Tosi, M.~P.} (1999).
\newblock Phys. Lett. A {\bf 260}, 86.

\bibitem[Chiofalo and Tosi, 2001]{Chiofalo:01}
\authorfont{Chiofalo, M.~L. and Tosi, M.~P.} (2001).
\newblock Europhys. Lett. {\bf 56}, 326.

\bibitem[Chu, 1998]{Chu:98}
\authorfont{Chu, S.} (1998).
\newblock Rev. Mod. Phys. {\bf 70}, 685.

\bibitem[Cohen-Tannoudji, 1998]{CohenTannoudji:98}
\authorfont{Cohen-Tannoudji, C.~N.} (1998).
\newblock Rev. Mod. Phys. {\bf 70}, 707.

\bibitem[Dalfovo {\em et~al.}, 1999]{Dalfovo:99}
\authorfont{Dalfovo, F., Giorgini, S., Pitaevskii, L.~P., and Stringari, S.}
  (1999).
\newblock Rev. Mod. Phys. {\bf 71}, 463.

\bibitem[{de Gennes}, 1966]{deGennes:66}
\authorfont{{de Gennes}, P.~G.}
\newblock {\em Superconductivity of Metals and Alloys} (W. A. Benjamin, New
  York, 1966).

\bibitem[{DeMarco} {\em et~al.}, 1998]{DeMarco:98}
\authorfont{{DeMarco}, B., Bohn, J.~L., {Burke, Jr.}, J.~P., Holland, M., and
  Jin, D.~S.} (1998).
\newblock Phys. Rev. Lett. {\bf 82}, 4208.

\bibitem[DeMarco and Jin, 1999]{DeMarco:99}
\authorfont{DeMarco, B. and Jin, D.~S.} (1999).
\newblock Science {\bf 285}, 1703.

\bibitem[Deng {\em et~al.}, 1999]{Deng:99}
\authorfont{Deng, L., Hagley, E.~W., Wen, J., Trippenbach, M., Band, Y.,
  Julienne, P.~S., Simsarian, J.~E., Helmerson, K., Rolston, S.~L., and
  Phillips, W.~D.} (1999).
\newblock Nature (London) {\bf 398}, 218.

\bibitem[Denschlag {\em et~al.}, 2000]{Denschlag:00}
\authorfont{Denschlag, J., Simsarian, J.~E., Feder, D.~L., Clark, C.~W.,
  Collins, L.~A., Cubizolles, J., Deng, L., Hagley, E.~W., Helmerson, K.,
  Reinhardt, W.~P., Rolston, S.~L., Schneider, B.~I., and Phillips, W.~D.}
  (2000).
\newblock Science {\bf 287}, 97.

\bibitem[Denschlag {\em et~al.}, 2002]{Denschlag:02}
\authorfont{Denschlag, J.~H., Simsarian, J.~E., Haffner, H., {McKenzie}, C.,
  Browaeys, A., Cho, D., Helmerson, K., Rolston, S.~L., and Phillips, W.~D.}
  (2002).
\newblock J. Phys. B: At. Mol. Opt. Phys. {\bf 35}, 3095.

\bibitem[Eiermann {\em et~al.}, 2003]{Eiermann:03}
\authorfont{Eiermann, B., Treutlein, P., Anker, T., Albiez, M., Taglieber, M.,
  Marzlin, K., and Oberthaler, M.~K.} (2003).
\newblock Phys. Rev. Lett. {\bf 91}, 060402.

\bibitem[Esteve {\em et~al.}, 2004]{Esteve:04}
\authorfont{Esteve, J., Aussibal, C., Schumm, T., Figl, C., Mailly, D.,
  Bouchoule, I., Westbrook, C., and Aspect, A.} (2004).
\newblock Phys. Rev. A {\bf 70}, 043629.

\bibitem[Fallani {\em et~al.}, 2003]{Fallani:03}
\authorfont{Fallani, L., Cataliotti, F.~S., Catani, J., Fort, C., Modugno, M.,
  Zawada, M., and Inguscio, M.} (2003).
\newblock Phys. Rev. Lett. {\bf 91}, 240405.

\bibitem[Fano, 1935]{Fano:35}
\authorfont{Fano, U.} (1935).
\newblock Nuovo Cimento {\bf 12}, 156.

\bibitem[Fazio and {van der Zant}, 2001]{Fazio:01}
\authorfont{Fazio, R. and {van der Zant}, H.} (2001).
\newblock Phys. Rep. {\bf 355}, 235.

\bibitem[Feshbach, 1958]{Feshbach:58}
\authorfont{Feshbach, H.} (1958).
\newblock Ann. Phys. (N.Y.) {\bf 5}, 357.

\bibitem[Folman {\em et~al.}, 2000]{Folman:00}
\authorfont{Folman, R., Kr\"uger, P., Cassettari, D., Hessmo, B., Maier, T.,
  and Schmiedmayer, J.} (2000).
\newblock Phys. Rev. Lett. {\bf 84}, 4749.

\bibitem[Fort {\em et~al.}, 2001]{Fort:01}
\authorfont{Fort, C., Maddaloni, P., Minardi, F., Modugno, M., and Inguscio,
  M.} (2001).
\newblock Optics Lett. {\bf 26}, 1039.

\bibitem[Fort {\em et~al.}, 2000]{Fort:00}
\authorfont{Fort, C., Prevedelli, M., Minardi, F., Cataliotti, F.~S., Ricci,
  L., Tino, G.~M., and Inguscio, M.} (2000).
\newblock Europhys. Lett. {\bf 49}, 8.

\bibitem[Fort\'agh {\em et~al.}, 2002]{Fortagh:02}
\authorfont{Fort\'agh, J., Ott, H., Kraft, S., G\"unther, A., and Zimmermann,
  C.} (2002).
\newblock Phys. Rev. A {\bf 66}, 041604(R).

\bibitem[Garc\'\i{}a-Ripoll and P\'erez-Garc\'\i{}a, 2001]{Garcia-Ripoll:01}
\authorfont{Garc\'\i{}a-Ripoll, J.~J. and P\'erez-Garc\'\i{}a, V.~M.} (2001).
\newblock Phys. Rev. A {\bf 63}, 041603.

\bibitem[Giovanazzi {\em et~al.}, 2002]{Giovanazzi:02}
\authorfont{Giovanazzi, S., G\"orlitz, A., and Pfau, T.} (2002).
\newblock Phys. Rev. Lett. {\bf 89}, 130401.

\bibitem[Greiner {\em et~al.}, 2002a]{Greiner:02}
\authorfont{Greiner, M., Mandel, O., Esslinger, T., H\"ansch, T.~W., and Bloch,
  I.} (2002a).
\newblock Nature (London) {\bf 415}, 39.

\bibitem[Greiner {\em et~al.}, 2002b]{Greiner:02a}
\authorfont{Greiner, M., Mandel, O., H\"ansch, T.~W., and Bloch, I.} (2002b).
\newblock Nature (London) {\bf 419}, 51.

\bibitem[Greiner {\em et~al.}, 2003]{Greiner:03}
\authorfont{Greiner, M., Regal, C.~A., and Jin, D.~S.} (2003).
\newblock Nature (London) {\bf 426}, 537.

\bibitem[Gross, 1961]{Gross:61}
\authorfont{Gross, E.~P.} (1961).
\newblock Nuovo Cimento {\bf 20}, 454.

\bibitem[Gross, 1963]{Gross:63}
\authorfont{Gross, E.~P.} (1963).
\newblock J. Math. Phys. {\bf 4}, 195.

\bibitem[Gygi and Schl\"uter, 1991]{Gygi:91}
\authorfont{Gygi, F. and Schl\"uter, M.} (1991).
\newblock Phys. Rev. B {\bf 43}, 7609.

\bibitem[Hadzibabic {\em et~al.}, 2002]{Hadzibabic:02}
\authorfont{Hadzibabic, Z., Stan, C.~A., Dieckmann, K., Gupta, S., Zwierlein,
  M.~W., G\"orlitz, A., and Ketterle, W.} (2002).
\newblock Phys. Rev. Lett. {\bf 88}, 160401.

\bibitem[H\"ansel {\em et~al.}, 2001a]{Haensel:01b}
\authorfont{H\"ansel, W., Hommelhoff, P., H\"ansch, T.~W., and Reichel, J.}
  (2001a).
\newblock Nature (London) {\bf 413}, 498.

\bibitem[H\"ansel {\em et~al.}, 2001b]{Haensel:01}
\authorfont{H\"ansel, W., Reichel, J., Hommelhoff, P., and H\"ansch, T.~W.}
  (2001b).
\newblock Phys. Rev. Lett. {\bf 86}, 608.

\bibitem[H\"ansel {\em et~al.}, 2001c]{Haensel:01a}
\authorfont{H\"ansel, W., Reichel, J., Hommelhoff, P., and H\"ansch, T.~W.}
  (2001c).
\newblock Phys. Rev. A {\bf 64}, 063607.

\bibitem[Harber {\em et~al.}, 2003]{Harber:03}
\authorfont{Harber, D.~M., {McGuirk}, J.~M., Obrecht, J.~M., and Cornell,
  E.~A.} (2003).
\newblock J. Low Temp. Phys. {\bf 133}, 229.

\bibitem[Hensler {\em et~al.}, 2003]{Hensler:03}
\authorfont{Hensler, S., Werner, J., Griesmaier, A., Schmidt, P.~O., G\"orlitz,
  A., Pfau, T., Giovanazzi, S., and {Rz\c{a}\.zewski}, K.} (2003).
\newblock Appl. Phys. B {\bf 77}, 765.

\bibitem[Hilligs{\o} {\em et~al.}, 2002]{Hilligsoe:02}
\authorfont{Hilligs{\o}, K.~M., Oberthaler, M.~K., and Marzlin, K.} (2002).
\newblock Phys. Rev. A {\bf 66}, 063605.

\bibitem[Howard {\em et~al.}, 2003]{Howard:03}
\authorfont{Howard, I.~A., Komarov, I.~V., March, N.~H., and Nieto, L.~M.}
  (2003).
\newblock J. Phys. A: Math. Gen. {\bf 36}, 4757.

\bibitem[Inguscio, 2003]{Inguscio:03}
\authorfont{Inguscio, M.} (2003).
\newblock Science {\bf 300}, 1671.

\bibitem[Inguscio {\em et~al.}, 1998]{Varenna:98}
\authorfont{Inguscio, M., Stringari, S., and Wieman, C.~E.} editors.
\newblock {\em Proceedings of the CXL International School of Physics {``E.
  Fermi''} on {``Bose-Einstein Condensation in Atomic Gases''}} (IOS,
  Amsterdam, 1998).

\bibitem[Inouye {\em et~al.}, 1998]{Inouye:98}
\authorfont{Inouye, S., Andrews, M.~R., Stenger, J., Miesner, H., {D. M.
  Stamper-Kurn}, and Ketterle, W.} (1998).
\newblock Nature (London) {\bf 392}, 151.

\bibitem[Inouye {\em et~al.}, 1999]{Inouye:99}
\authorfont{Inouye, S., Pfau, T., Gupta, S., Chikkatur, A.~P., Gorlitz, A.,
  Pritchard, D.~E., and Ketterle, W.} (1999).
\newblock Nature (London) {\bf 402}, 641.

\bibitem[Jaksch {\em et~al.}, 1998]{Jaksch:98}
\authorfont{Jaksch, D., Bruder, C., Cirac, J.~I., Gardiner, C.~W., and Zoller,
  P.} (1998).
\newblock Phys. Rev. Lett. {\bf 81}, 3108.

\bibitem[Jochim {\em et~al.}, 2003]{Jochim:03}
\authorfont{Jochim, S., Bartenstein, M., Altmeyer, A., Hendl, G., Riedl, S.,
  Chin, C., {Hecker Denschlag}, J., and Grimm, R.} (2003).
\newblock Science {\bf 302}, 2101.

\bibitem[Jones {\em et~al.}, 2003]{Jones:03}
\authorfont{Jones, M. P.~A., Vale, C.~J., Sahagun, D., Hall, B.~V., and Hinds,
  E.~A.} (2003).
\newblock Phys. Rev. Lett. {\bf 91}, 080401.

\bibitem[Jones and March, 1986]{JM1}
\authorfont{Jones, W. and March, N.~H.}
\newblock {\em Theoretical Solid-State Physics. Perfect Lattices in
  Equilibrium}, volume~1 (Dover, London, 1986).

\bibitem[Khaykovich {\em et~al.}, 2002]{Khaykovich:02}
\authorfont{Khaykovich, L., Schreck, F., Ferrari, G., Bourdel, T., Cubizolles,
  J., Carr, L.~D., Castin, Y., and Salomon, C.} (2002).
\newblock Science {\bf 296}, 1290.

\bibitem[Konotop and Salerno, 2002]{Konotop:02}
\authorfont{Konotop, V.~V. and Salerno, M.} (2002).
\newblock Phys. Rev. A {\bf 65}, 021602.

\bibitem[Kozuma {\em et~al.}, 1999]{Kozuma:99}
\authorfont{Kozuma, M., Suzuki, Y., Torii, Y., Sugiura, T., Kuga, T., Hagley,
  E.~W., and Deng, L.} (1999).
\newblock Science {\bf 286}, 2309.

\bibitem[Leggett, 2001]{Leggett:01}
\authorfont{Leggett, A.~J.} (2001).
\newblock Rev. Mod. Phys. {\bf 73}, 307.

\bibitem[Leggett, 2003]{Leggett:03}
\authorfont{Leggett, A.~J.} (2003).
\newblock New J. Phys. {\bf 5}, 103.1.

\bibitem[Lenz {\em et~al.}, 1994]{Lenz:94}
\authorfont{Lenz, G., Meystre, P., and Wright, E.~M.} (1994).
\newblock Phys. Rev. A {\bf 50}, 1681.

\bibitem[Lieb and Yngvason, 1998]{Lieb:98}
\authorfont{Lieb, E.~H. and Yngvason, J.} (1998).
\newblock Phys. Rev. Lett. {\bf 80}, 2504.

\bibitem[Lin {\em et~al.}, 2004]{Lin:04}
\authorfont{Lin, Y., Teper, I., Chin, C., and Vuleti\'c, V.} (2004).
\newblock Phys. Rev. Lett. {\bf 92}, 050404.

\bibitem[Makhlin {\em et~al.}, 2001]{Makhlin:01}
\authorfont{Makhlin, Y., Sch\"on, G., and Shnirman, A.} (2001).
\newblock Rev. Mod. Phys. {\bf 73}, 357.

\bibitem[Mancini {\em et~al.}, 2004]{Mancini:04}
\authorfont{Mancini, M.~W., Telles, G.~D., Caires, A. R.~L., Bagnato, V.~S.,
  and Marcassa, L.~G.} (2004).
\newblock Phys. Rev. Lett. {\bf 92}, 133203.

\bibitem[Mandel {\em et~al.}, 2003]{Mandel:03}
\authorfont{Mandel, O., Greiner, M., Widera, A., Rom, T., H\"ansch, T.~W., and
  Bloch, I.} (2003).
\newblock Nature (London) {\bf 425}, 937.

\bibitem[March and Murray, 1960]{March:60}
\authorfont{March, N.~H. and Murray, A.~M.} (1960).
\newblock Phys. Rev. {\bf 120}, 830.

\bibitem[March and Murray, 1961]{March:61}
\authorfont{March, N.~H. and Murray, A.~M.} (1961).
\newblock Proc. Roy. Soc. A {\bf 261}, 119.

\bibitem[March {\em et~al.}, 1995]{March:95}
\authorfont{March, N.~H., Young, W.~H., and Sampanthar, S.}
\newblock {\em The Many-Body Problem in Quantum Mechanics} (Dover, New York,
  1995).

\bibitem[Massignan and Modugno, 2003]{Massignan:03}
\authorfont{Massignan, P. and Modugno, M.} (2003).
\newblock Phys. Rev. A {\bf 67}, 023614.

\bibitem[Mewes {\em et~al.}, 1997]{Mewes:97}
\authorfont{Mewes, M., Andrews, M.~R., Kurn, D.~M., Durfee, D.~S., Townsend,
  C.~G., and Ketterle, W.} (1997).
\newblock Phys. Rev. Lett. {\bf 78}, 582.

\bibitem[Minguzzi {\em et~al.}, 2004]{Minguzzi:04}
\authorfont{Minguzzi, A., Succi, S., Toschi, F., Tosi, M.~P., and Vignolo, P.}
  (2004).
\newblock Phys. Rep. {\bf 395}, 223.

\bibitem[Nozi{\`e}res and {Schmitt-Rink}, 1985]{Nozieres:85}
\authorfont{Nozi{\`e}res, P. and {Schmitt-Rink}, S.} (1985).
\newblock J. Low Temp. Phys. {\bf 59}, 195.

\bibitem[{O'Dell} {\em et~al.}, 2004]{ODell:04}
\authorfont{{O'Dell}, D. H.~J., Giovanazzi, S., and Eberlein, C.} (2004).
\newblock Phys. Rev. Lett. {\bf 92}, 250401.

\bibitem[Oliveira {\em et~al.}, 1988]{Oliveira:88}
\authorfont{Oliveira, L.~N., Gross, E. K.~U., and Kohn, W.} (1988).
\newblock Phys. Rev. Lett. {\bf 60}, 2430.

\bibitem[Ott {\em et~al.}, 2001]{Ott:01}
\authorfont{Ott, H., Fortagh, J., Schlotterbeck, G., Grossmann, A., and
  Zimmermann, C.} (2001).
\newblock Phys. Rev. Lett. {\bf 87}, 230401.

\bibitem[Paredes {\em et~al.}, 2004]{Paredes:04}
\authorfont{Paredes, B., Widera, A., Murg, V., Mandel, O., F\"olling, S.,
  Cirac, I., Shlyapnikov, G.~V., H\"ansch, T.~W., and Bloch, I.} (2004).
\newblock Nature (London) {\bf 429}, 277.

\bibitem[Parkins and Walls, 1998]{Parkins:98}
\authorfont{Parkins, A.~S. and Walls, D.~F.} (1998).
\newblock Phys. Rep. {\bf 303}, 1.

\bibitem[Pedri {\em et~al.}, 2001]{Pedri:01}
\authorfont{Pedri, P., Pitaevskii, L., Stringari, S., Fort, C., Burger, S.,
  Cataliotti, F.~S., Maddaloni, P., Minardi, F., and Inguscio, M.} (2001).
\newblock Phys. Rev. Lett. {\bf 87}, 220401.

\bibitem[Phillips, 1998]{Phillips:98}
\authorfont{Phillips, W.~D.} (1998).
\newblock Rev. Mod. Phys. {\bf 70}, 721.

\bibitem[Pieri and Strinati, 2000]{Pieri:00}
\authorfont{Pieri, P. and Strinati, G.~C.} (2000).
\newblock Phys. Rev. B {\bf 61}, 15370.

\bibitem[Pieri and Strinati, 2003]{Pieri:03}
\authorfont{Pieri, P. and Strinati, G.~C.} (2003).
\newblock Phys. Rev. Lett. {\bf 91}, 030401.

\bibitem[Pitaevskii, 1961]{Pitaevskii:61}
\authorfont{Pitaevskii, L.~P.} (1961).
\newblock Zh. Eksp. Teor. Fiz. {\bf 40}, 646.
\newblock [Sov. Phys. JETP {\bf 13}, 451 (1961)].

\bibitem[Raab {\em et~al.}, 1987]{Raab:87}
\authorfont{Raab, E.~L., Prentiss, M., Cable, A., Chu, S., and Pritchard,
  D.~E.} (1987).
\newblock Phys. Rev. Lett. {\bf 59}, 2631.

\bibitem[Randeria, 1995]{Randeria:95}
\authorfont{Randeria, M.} (1995).
\newblock Crossover from {BCS} Theory to {Bose-Einstein} Condensation.
\newblock In {\em {Bose} {Einstein} Condensation}, edited by Griffin, A.,
  Snoke, D., and Stringari, S., p. 355 (Cambridge University Press, Cambridge,
  1995).

\bibitem[Roati {\em et~al.}, 2002]{Roati:02}
\authorfont{Roati, G., Riboli, F., Modugno, G., and Inguscio, M.} (2002).
\newblock Phys. Rev. Lett. {\bf 89}, 150403.

\bibitem[Sachdev, 1999]{Sachdev:99}
\authorfont{Sachdev, S.}
\newblock {\em Quantum Phase Transitions} (Cambridge University Press,
  Cambdridge, 1999).

\bibitem[Schmidt {\em et~al.}, 2003]{Schmidt:03}
\authorfont{Schmidt, P.~O., Hensler, S., Werner, J., Griesmaier, A., G\"orlitz,
  A., Pfau, T., and Simoni, A.} (2003).
\newblock Phys. Rev. Lett. {\bf 91}, 193201.

\bibitem[Schreck {\em et~al.}, 2001]{Schreck:01}
\authorfont{Schreck, F., Khaykovich, L., Corwin, K.~L., Ferrari, G., Bourdel,
  T., Cubizolles, J., and Salomon, C.} (2001).
\newblock Phys. Rev. Lett. {\bf 87}, 080403.

\bibitem[Schrieffer, 1964]{Schrieffer:64}
\authorfont{Schrieffer, J.~R.}
\newblock {\em Theory of Superconductivity} (W. A. Benjamin, New York, 1964).

\bibitem[Sondheimer and Wilson, 1951]{Sondheimer:51}
\authorfont{Sondheimer, E.~H. and Wilson, A.~H.} (1951).
\newblock Proc. Roy. Soc. A {\bf 210}, 173.

\bibitem[Stoddart {\em et~al.}, 1968]{Stoddart:68}
\authorfont{Stoddart, J.~C., Hilton, D., and March, N.~H.} (1968).
\newblock Proc. Roy. Soc. A {\bf 304}, 99.

\bibitem[Stringari, 1996]{Stringari:96}
\authorfont{Stringari, S.} (1996).
\newblock Phys. Rev. Lett. {\bf 77}, 2360.

\bibitem[Tilley and Tilley, 1990]{Tilley:90}
\authorfont{Tilley, D.~R. and Tilley, J.}
\newblock {\em Superfluidity and Superconductivity} (Hilger, New York, 1990).

\bibitem[Timmermans {\em et~al.}, 2001]{Timmermans:01}
\authorfont{Timmermans, E., Furuya, K., Milonni, P.~W., and Kerman, A.~K.}
  (2001).
\newblock Phys. Lett. A {\bf 285}, 228.

\bibitem[Treutlein {\em et~al.}, 2004]{Treutlein:04}
\authorfont{Treutlein, P., Hommelhoff, P., Steinmetz, T., H\"ansch, T., and
  Reichel, J.} (2004).
\newblock Phys. Rev. Lett. {\bf 92}, 203005.

\bibitem[Trombettoni and Smerzi, 2001]{Trombettoni:01}
\authorfont{Trombettoni, A. and Smerzi, A.} (2001).
\newblock Phys. Rev. Lett. {\bf 86}, 2353.

\bibitem[Truscott {\em et~al.}, 2001]{Truscott:01}
\authorfont{Truscott, A.~G., Strecker, K.~E., {McAlexander}, W.~I., Partridge,
  G.~B., and Hulet, R.~G.} (2001).
\newblock Science {\bf 291}, 2570.

\bibitem[Weinstein {\em et~al.}, 1998]{Weinstein:98}
\authorfont{Weinstein, J.~D., {deCarvalho}, R., Guillet, T., Friedrich, B., and
  Doyle, J.~M.} (1998).
\newblock Nature (London) {\bf 395}, 148.

\bibitem[Weiss and Eckardt, 2004]{Weiss:04}
\authorfont{Weiss, C. and Eckardt, A.} (2004).
\newblock Europhys. Lett. {\bf 68}, 8.

\bibitem[Wierzbowska and Krogh, 2005]{Wierzbowska:04}
\authorfont{Wierzbowska, M. and Krogh, J.~W.} (2005).
\newblock Phys. Rev. B {\bf 71}, 014509.

\bibitem[Zwierlein {\em et~al.}, 2003]{Zwierlein:03}
\authorfont{Zwierlein, M.~W., Stan, C.~A., Schunck, C.~H., Raupach, S. M.~F.,
  Gupta, S., Hadzibabic, Z., and Ketterle, W.} (2003).
\newblock Phys. Rev. Lett. {\bf 91}, 250401.

\end{thebibliography}

\end{document}